\documentclass[12pt]{amsart}

\usepackage[english]{babel}

\usepackage[letterpaper,top=2cm,bottom=2cm,left=3cm,right=3cm,marginparwidth=1.75cm]{geometry}

\usepackage{hhline}
\usepackage{booktabs}
\usepackage{amsmath, amssymb}
\usepackage{algorithm}
\usepackage{algorithmic}
\usepackage{subcaption} 

\DeclareMathOperator*{\argmin}{arg\,min}
\usepackage{graphicx}
\usepackage{xcolor}
\usepackage[colorlinks=true, allcolors=blue]{hyperref}

\title[Sample continuation in Bayesian hierarchical model]{Sample continuation in Bayesian hierarchical model via variational inference}
\author{Yucong Liu}
\address{School of Mathematics, Georgia Institute of Technology, Atlanta, GA 30332}
\email{yucongliu@gatech.edu}

\author{Zilai Si}
\address{Department of Industrial Engineering and Management Sciences, Northwestern University, Evanston, IL 60208}
\email{zilaisi2028@u.northwestern.edu}

\author{Alexander Strang}
\address{Department of Statistics, University of California Berkeley, Berkeley, CA 94720} \email{alexstrang@berkeley.edu}

\begin{document}
\maketitle

\begin{abstract}
% This is an example SIAM \LaTeX\ article. This can be used as a
% template for new articles.  Abstracts must be able to stand alone
% and so cannot contain citations to the paper's references,
% equations, etc.  An abstract must consist of a single paragraph and
% be concise. Because of online formatting, abstracts must appear as
% plain as possible. Any equations should be inline.
Posterior distributions arising in ill-posed Bayesian inverse problems are often both analytically intractable and highly sensitive to parameters of the chosen prior family. We aim to understand the sensitivity of intractable posterior distributions to changes in prior assumptions by tracking how a sample representation of the posterior changes as the prior parameters change.
This enables sensitivity analysis for small perturbations in the prior, providing insights into the robustness of the posterior estimates under minor changes in assumptions. It also allows solution continuation when dealing with significant alterations in prior beliefs, facilitating a comprehensive understanding of how large shifts in assumptions affect the posterior distribution. We focus on a class of non-conjugate hierarchical models tailored to encourage sparsity in linear inverse problems. The specific hierarchical model of interest is chosen since it is parameterized by a small number of shape parameters, and includes most classical sparsity promoting priors as special cases.  As the shape parameters change, the posterior can transition continuously from a tractable unimodal distribution to an intractable multimodal distribution.
To track the change in the posterior, we adopt particle based variational inference methods, specifically
Stein Variational Gradient Descent (SVGD). SVGD iteratively updates a set of samples to minimize the KL-divergence away from a desired target distribution. We augment SVGD by Birth-Death sampling, which can efficiently exchange mass between separated modes, while simultaneously optimizing the kernel bandwidth used to derive the SVGD update. This method enables the discovery of new modes by tracing the modes as they branch out of a simpler, unimodal posterior, derived within the same family of priors. Experimental results demonstrate that this approach can sample from multimodal posteriors more comprehensively than existing methods and allows efficient sensitivity analysis.
\end{abstract}

% % REQUIRED
% \begin{keywords}
% bayesian hierarchical model, variational inference, distribution evolution
% \end{keywords}

% % REQUIRED
% \begin{MSCcodes}
% 62F15, 65C30
% \end{MSCcodes}

\section{Introduction}
%\red{background on inverse problem, why linear inverse problem}

In an inverse problem, a user aims to recover an unknown signal $ x \in \mathbb{R}^n $ from noisy observations of a transformation of $x$, $ b \in \mathbb{R}^m $. A linear inverse problem can be defined:
\begin{equation}\label{eqn:linear_inverse_prob}
    b = Ax + \epsilon,
\end{equation}
where $ A \in \mathbb{R}^{m \times n} $ is a known measurement matrix, and $ \epsilon $ represents noise. Linear inverse problems have a great impact on multiple areas including geophysics \cite{zhdanov2002geophysical, oldenburg1984introduction}, electromagnetism \cite{de1998critical,phillips2002systematic},  and image processing \cite{ribes2008linear, imageinverse}, especially medical imaging \cite{medical, LinearProblemApplication, MathMedical}.

%\red{sparsity}

Most inverse problems of interest are ill-posed \cite{candes2006stable, donoho2005stable, figueiredo2007gradient}, so, are intractable without assumptions that restrict the domain of possible solutions, or that preferentially select among plausible $x$. %A classical approach to address this challenge is to reformulate the problem by introducing a penalized functional whose minimizer provides a stable and interpretable solution. This leads to a regularized objective function of the form
%
%\begin{equation}
%F_{p}(x, \lambda) = \|b - Ax\|^2 + \lambda \| x\|_p^p,
%\end{equation}
%
%where the data fidelity term, $\|b - Ax\|^2$, ensures that $A x$ is close to the observed signal, $b$, and $\|x\|_p^p$ is a regularization term designed to promote sparsity in the proposed solution. The regularization parameter, $\lambda$,balances the trade-off between the data fidelity and the penalty term, thereby determining the degree of shrinkage demanded in the solution. The parameter $p \in [0,2)$, affects the shape of the sparsity promoting term. Traditionally $p = 1$ is used as a tractable relaxation of the $p = 0$ penalty, which counts the number of nonzero components in the proposed solution. 
% for example, iteratively reweighted least squares (IRLS) algorithms \cite{FOCUSS, daubechies2010iteratively} has been proposed for $1\leq p < 2$.
%
%
Bayesian inference incorporates prior information about the unknown to bias inference towards solutions that are more likely before observing the data. In a Bayesian setting the unknown signal $ x $ is modeled as a random variable. The prior distribution, ${ \pi_{\text{prior}}(x \mid \psi) }$, encodes essential information about $ x $, such as sparsity, where $ \psi $ denotes the associated parameters. With this prior knowledge, the posterior distribution of $ x $, conditioned on the observation $ b $ and hyperparameters $ \psi $, provides a probabilistic solution to the linear inverse problem described in Equation \ref{eqn:linear_inverse_prob}. Typically, posterior means or modes are returned as point estimators, and samples from the posterior distribution are used to quantify the uncertainty in those estimates, often through the construction of credible intervals.

%\red{Bayesian model}

Suppose $ \epsilon \sim \mathcal{N}(0,\Sigma)$. Then, the likelihood of observing the data $ b $, given $ x $, follows a normal distribution:
\begin{equation} \label{eqn: likelihood}
    \pi(b|x) = \frac{1}{\sqrt{(2 \pi)^{m} |\det(\Sigma)|}} \exp \left(-\frac{1}{2}(Ax-b)^\top \Sigma^{-1} (Ax - b) \right).
\end{equation}
The posterior distribution of $ x $ can then be expressed:
\begin{equation}
\pi_{\text{post}}(x \mid b,\psi) \propto \pi(b \mid x) \pi_{\text{prior}}(x \mid \psi),
\end{equation}
where $ \pi_{\text{prior}}(x \mid \psi) $ encodes probabilistic assumptions about the unknown. These are fixed by the form of the prior, and its parameters, $\psi$. %Then maximum a posteriori estimate is also chosen to be the Bayesian solution.

%\red{Hyperparameter changes, sensitivity}

In practice, the underlying parameters $ \psi $ are rarely known with great precision, and are often chosen subjectively to satisfy downstream desiderata like computational tractability.  Since the inferential results depend on the choice of prior, it is essential to study the model's sensitivity to changes in the prior. In this paper, we attempt to track the posterior as the parameters $\psi$ vary smoothly. To model these changes, we construct continuous paths, $\psi(t)$, in parameter space, where $t$ is an artificial time variable. Then, the posterior distribution becomes time-dependent:
\begin{equation}
\pi_{\text{post}}(x \mid b, \psi(t)) \propto \pi(b \mid x) \, \pi_{\text{prior}}(x \mid \psi(t)),
\end{equation}
where $\pi_{\text{prior}}(x \mid \psi(t))$ encapsulates prior knowledge about $x$ that adapts according to $\psi(t)$. The sensitivity of the maximum a posteriori (MAP) estimate has been explored in \cite{si2024path}, but a broader perspective on the entire posterior distribution is needed for a more comprehensive understanding. The lack of an explicit form for the posterior distribution prevents direct study. To overcome this challenge, we rely on a sample representation to approximate and analyze the distribution. This approach necessitates a robust sampling method whose sample approximations to the posterior remain representative, even as it evolves.

%\red{literature review}

A variety of methods exist for tracking sample representations of distributions as they change. For example, Sequential Monte Carlo (SMC) methods, also known as particle filters, are widely used for handling sequential simulations in discrete-time dynamic models, particularly in nonlinear and non-Gaussian settings \cite{doucet2000sequential, chopin2002sequential, chopin2020introduction}. However, SMC methods typically assume an underlying hidden Markov process with a known transition distribution $p(x_t \mid x_{t-1})$, which imposes a stronger assumption than is required in our setting.

We adopt variational Bayesian inference instead. Variational methods perform approximate posterior inference by replacing an intractable posterior with its closest approximation restricted to a tractable family \cite{blei2017variational}. Variational methods offer several advantages over traditional sampling methods in a continuation setting, most notably, the variational solutions may be expressed as functions of the posterior, and thus the prior, via the implicit function theorem.

Normalizing flows offer an intuitive variational framework for tracking continuously changing posteriors when the initial posterior is tractable. A normalizing flow \cite{NormalizingFlows} describes the transformation of a tractable probability density through a sequence of invertible mappings. The flow is chosen so that transformed samples from the original distribution approximate the desired posterior distribution. The full map is constructed iteratively by composing a sequence of component transformations that are each close to an identity map. Applying the transformations sequentially to an initial sample produces a discrete time dynamical system. The motion of a sample under this system is the flow. Each component transformation is usually selected to follow a functional gradient of the discrepancy between the variational density and the target in a selected norm, and under a chosen scheme for approximating the gradient given samples from the variational density \cite{wang2021particle}. Classically, each component mapping is a neural network. Training the component mappings can be computationally intensive and practical implementation requires expertise, for example, in choosing the network architecture \cite{liu2016stein}. 
%
%If we consider the ordinary differential equations governing the evolution of a distribution’s density, we encounter the classic Fokker-Planck equations \cite{risken1996fokker}. 
%
 
 %The Jordan-Kinderleherer-Otto (JKO) scheme \cite{JKO}, also known as the Wasserstein Gradient Flow, was developed to solve this variational problem by identifying the distribution’s path of steepest descent with respect to the Wasserstein metric. This foundational approach has inspired numerous recent works \cite{lambert2022variational, pmlr-v162-fan22d, yan2024learning}. However, in practice, computing the Wasserstein distance between two probability densities is challenging, especially in high-dimensional spaces. Additionally, each step in the JKO scheme requires solving an optimization problem, which can be computationally expensive. 

 These limitations motivate particle based alternatives. Particle based alternatives substitute the construction of an explicit mapping, as in \cite{NormalizingFlows}, with the time evolution of a fixed set of initial samples. Stein Variational Gradient Descent (SVGD) \cite{liu2016stein}, sets each component transformation equal to an approximation to the functional gradient of the KL-divergence within a unit ball of a reproducing kernel Hilbert space (RKHS). The descent direction is approximated at each stage using the set of samples. SVGD can efficiently approximate continuous changes in the posterior distributions via particle-based updates. SVGD is an implementation of the more general Energetic Variational Inference (EVI) framework, which replaces the RKHS norm with norms derived from optimal transport, and which allows for different kernelized approximation schemes for estimating the functional gradients \cite{wang2021particle}. 
 
 In this work, we adopt SVGD, then augment its sample updating process to address its limitations. Namely, SVGD cannot transfer mass between sufficiently well separated modes, even if the modes change in mass, and, SVGD produces sample estimates that depend on, and are biased by, the choice of a kernel and kernel bandwidth. In particular, we do not adopt the standard ``median trick" \cite{liu2016stein,wang2021particle}, which overestimates the bandwidth for multimodal or eccentric distributions. Instead, we optimize the bandwidth in parallel with the particle locations. 
 
%\red{why continuation}

This augmented SVGD algorithm can efficiently update a sample representation of a varying posterior. When the prior only changes slightly, then the augmented SVGD algorithm allows for sensitivity analysis of any posterior quantity that can be estimated using samples. Importantly, this allows sensitivity analysis for UQ products such as credible intervals. When the posterior changes significantly, then iterative application of the augmented SVGD allows sample continuation. Sample continuation is particularly useful if the desired prior produces a highly multimodal posterior, as is often the case for sparse Bayesian inference \cite{bayeslasso, seeger2007bayesian}.

\begin{figure}[h]
     \centering
     \includegraphics[width=.7\linewidth]{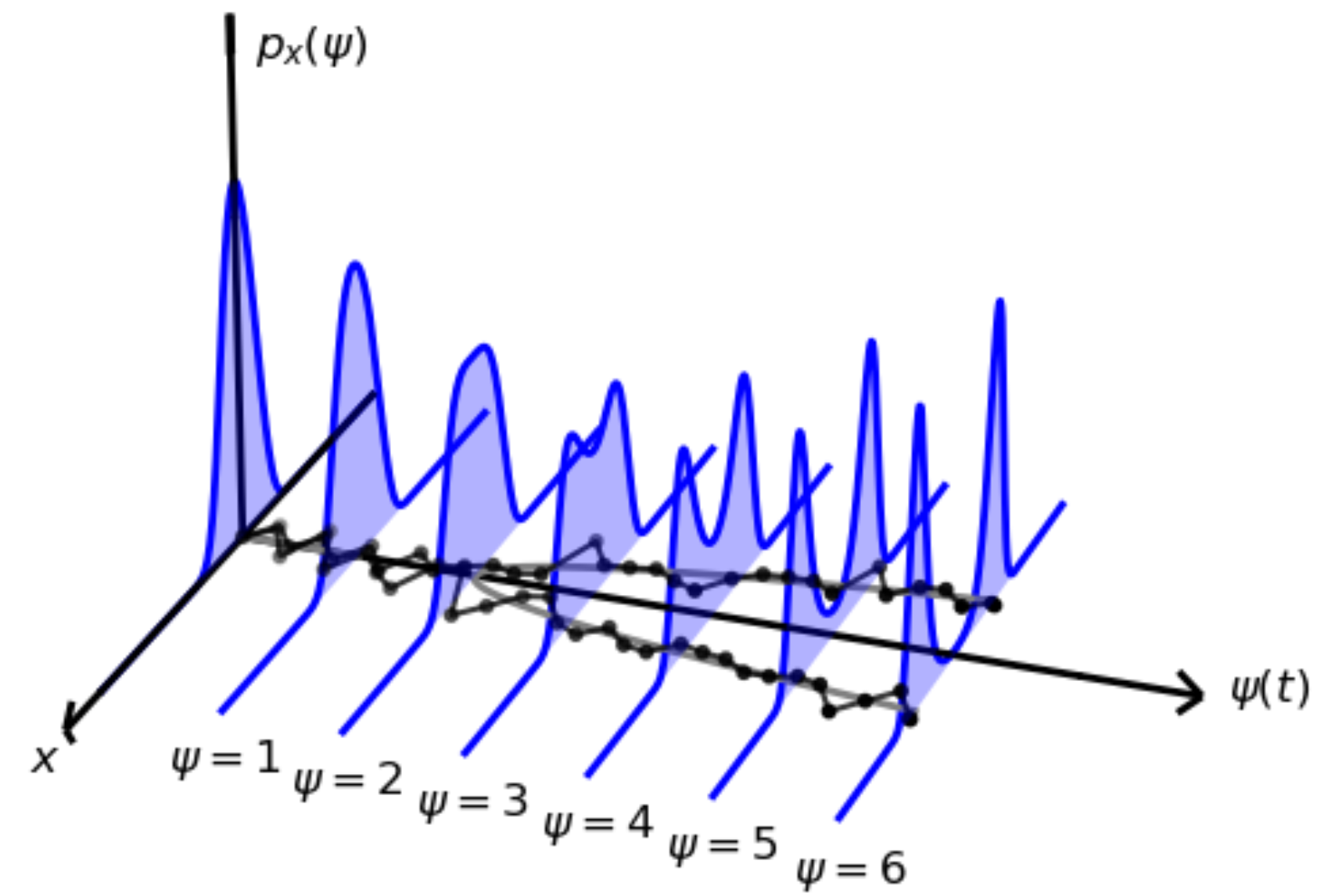}
     
    \caption{ A unimodal distribution transitions into a multimodal distribution as the parameter $\psi$ increases. The distribution is set to be $
p_{\psi}(x) \sim \exp \left( - \left( x^4 - (\psi - 2)x^2 \right) \right)
$. For $\psi \leq 2$, the distribution remains unimodal, but as 
$ \psi$ increases, it bifurcates into a multimodal form.
and breaks into a mutimodal distribnution as $\psi$ increase. The blue curve represents the probability density function for different values of $\psi$, while the two black curves on the $\psi-x$ plane depict the trajectories of two particles. Initially originating from the same point at the origin, these particles diverge and settle into distinct modes as the unimodal distribution splits.}
\label{fig:3path}
\end{figure}

Consider a sparsity-promoting prior that induces a highly multimodal posterior. Standard methods for sampling directly from highly multimodal posteriors are slow, and are not guaranteed to discover all modes. For example, Markov chain Monte Carlo (MCMC) algorithms struggle to approximate multimodal posteriors since the time needed to discover, and transition between, well separated modes is often excessively long \cite{yao2022stacking, MR2288719, feroz2013importance}. Moreover, mode discovery is directed by unreliable random re-initialization of a MAP estimation pipeline. Nevertheless, most sparsity promoting priors are members of a larger family of continuously related priors. If that family includes priors such that the posterior is unimodal, then there may exist paths through the prior parameter space that connect a multimodal and unimodal posterior distribution. Sample continuation tools use standard samplers to draw from the unimodal posterior, then update the samples to track changes in the posterior along the path linking the tractable unimodal posterior, and the intractable target posterior. Provided new modes are formed by a branching, or bifurcating process, then new modes are easily discovered as they first form, then reliably tracked as they separate, which is illustrated in Figure \ref{fig:3path}. As a result, sample continuation offers a simple procedure for finding the modes of a multimodal posterior if the modes branch out of a single mode under a related prior.

This evolving framework leverages the continuity of the process: if the distribution changes gradually, it is more efficient to update each sample incrementally as the distribution shifts, rather than resampling the entire system from scratch at each step. This approach reduces computational cost and improves convergence, as samples can adapt to the changing landscape without needing to explore the entire distribution repeatedly. %, making it particularly advantageous for handling complex, multi-modal distributions.

This paper is organized as follows: In Section \ref{sec:model} we introduce the Bayesian Hierarchical Model. Then we introduce Sample Continuation Methods in Section \ref{sec:method}, including SVGD in Section \ref{sec:svgd}, Bandwidth Update in Section \ref{sec:bandwidth}, a Birth-Death Process in Section \ref{sec:birthdeath} and a Back Tracking Technique in Section \ref{sec:backtrack}. Finally, our experiments, presented in Section \ref{sec:experiment}, demonstrate the effectiveness of our approach in capturing sample representations for multimodal distributions.

\section{Bayesian Hierarchical Modeling} \label{sec:model}
%\red{Define Bayesian Hierarchical Model}

In this work, we focus on a hierarchical model, in which we suppose that $x$ is drawn from a conditionally Gaussian prior distribution with variances $\theta$ that themselves follow a generalized gamma distribution. This Bayesian hierarchical model, was introduced in \cite{calvetti2008hypermodels,calvetti2009conditionally,calvetti2020sparse}, for electroencephalography (EEG) and magnetoencephalography (MEG) inverse problems in brain imaging \cite{calvetti2015hierarchical,calvetti2019brain}. Its hierarchical structure provides expressivity, the chosen scale mixture of Gaussians can express most standard sparsity-promoting or robust-inference priors \cite{andrews1974scale, choy1997hierarchical,ouyang2022robust}, while allowing efficient MAP estimation in large problems via coordinate ascent algorithms that exploit the simplicity of each component of the hierarchical model \cite{calvetti2009conditionally}. These produce an iteratively reweighted least-squares procedure with weight updates derived by maximizing the posterior over $\theta$ conditional on $x$ \cite{calvetti2019hierachical}.

Specifically, we assume that $\pi(x|\theta)$ is Gaussian distribution $\mathcal{N}(0,D_{\theta})$ where $D_{\theta}$ denotes a diagonal matrix with diagonal entries specified by the vector $\theta$. The variances $\theta \in \mathbb{R}^n$ are drawn independently from a generalized gamma distribution with parameters $r$, $\eta$ and $\vartheta_j$. Thus:
\begin{equation}
    \pi_{\text{hyper}}(\theta_j|r,\eta,\vartheta_j) =  \frac{|r|^n}{\Gamma(\beta)^n} \prod_{j=1}^n \frac{1}{\vartheta_j} \left( \frac{\theta_j}{\vartheta_j} \right)^{\eta + \frac{1}{2}} \exp \left(-\left( \frac{\theta_j}{\vartheta_j} \right)^r \right),
\end{equation}
where $\eta = r \beta - 3/2$ \cite{calvetti2009conditionally,calvetti2020sparse}. The parameters $r \in \mathbb{R} \setminus \{0\} $ and $\eta$ are shape parameters, chosen such that $\beta > 0$. The parameters $\vartheta_j > 0$ are scale parameters.

%\red{Define Posterior}

The complete posterior distribution is given by 
$$
\pi(x, \theta | b) \propto \pi(b | x) \pi_{\text{prior}}(x|\theta) \pi_{\text{hyper}}(\theta | r, \eta, \vartheta).
$$ 
Since the prior $ \pi_{\text{hyper}}(\theta | r, \eta, \vartheta) $ is non-conjugate, the resulting posterior distribution is not Gaussian and is only available up to an unknown normalizing factor. The negative logarithm of the posterior, also known as the Gibbs energy, is:
\begin{equation}
    \mathcal{G}(x, \theta) = \frac{1}{2} \|b - A x\|^2 + \frac{1}{2} \sum_{j=1}^{n} \frac{x_j^2}{\theta_j} - \eta \sum_{j=1}^{n} \log \frac{\theta_j}{\vartheta_j} + \sum_{j=1}^{n} \left( \frac{\theta_j}{\vartheta_j} \right)^r 
\end{equation}
up to an additive constant determined by normalization. 

In this Bayesian Hierarchical Model, parameters $r, \eta, \vartheta$ play a crucial role in shaping the posterior distribution of $x, \theta$. Studying the maximum a posterior estimation problem can provide some intuition for the influence of the parameters. If $\eta$ converges to zero, then the regularizer converges to the corresponding $\ell_p$ penalty, and the MAP estimation
problem reduces to the standard $\ell_p$ regularized least squares problem. Here, $p = \frac{2r}{1+r}$ \cite{calvetti2019hierachical}. Moreover, specific choices of $r$ and $\eta$ yield well-known distributions: for example, setting $r = -1$ results in a Student-t prior on $x$ when marginalizing over $\theta$; choosing $r = 1$, $\eta = -1/2$ leads to a Laplace prior; and for sufficiently large $r$and $\eta$, the prior approaches a normal distribution.  

Now, consider a scenario where the parameters $\psi(t) = [r(t),\eta(t),\vartheta(t)]$ evolve smoothly as a function of some fictitious time  
$t$. As these parameters trace a continuous path through the parameter space, the corresponding posterior distributions of $x$ and $\theta$ also shift accordingly. Our main task is to sample $x$ and $\theta$ following $\psi(t)$.

\section{Sample Continuation Methods}
\label{sec:method}
\subsection{Stein Variational Gradient Descent}
\label{sec:svgd}
% \red{Main Method: SVGD}

Variational inference procedures approximate a generic target distribution $\rho^*$ by minimizing a divergence between $\rho \in \mathcal{Q}$ and $\rho^*$, where $\mathcal{Q}$ represents a family of tractable distributions. Typically, variational inference procedures aim to minimize the Kullbach-Leibler (KL) divergence.

In the EVI framework \cite{wang2021particle}, the set $\mathcal{Q}$ consists of all distributions that can be obtained through smooth transformations of a tractable initial reference distribution. Specifically, the goal is to identify the optimal smooth, one-to-one mapping $z = \phi(x)$
that minimizes the KL divergence between the distribution of $z$, denoted as $\rho_\phi(z)$, and the target distribution $\rho^*(z)$, where $x$ is sampled from a tractable initial distribution $\rho(x)$.  Then, the density of $z$ is given by:
\begin{equation}
    \rho_{\phi}(z) = \rho(\phi^{-1}(z)) \left|\det \nabla_z \phi^{-1}(z) \right|.
\end{equation}

The KL-divergence with target distribution $\rho^{*}$ is: 
\begin{equation}
\begin{aligned}
    \text{KL}(\rho_{\phi} \parallel \rho^*) & = \int_z \rho_{\phi}(z) \log \frac{\rho_{\phi}(z)}{\rho^{*}(z)} dz \\
    & 
    =
    \int_x \rho(x) \left[\log \left(\frac{\rho(x)}{\left|\det \nabla_x \phi(x) \right|} \right) - \log(\rho^{*}(\phi(x))) \right] dx
\end{aligned}
\end{equation}
which is a functional of the smooth one-to-one map $\phi$. 

Generally there is no explicit solution for the optimal smooth transformation $\phi^{*}$ that minimizes the KL divergence:
$$
\phi^{*} = \argmin\{\text{KL}(\rho_{\phi} \parallel \rho^{*})\}
$$

Instead of solving for an optimal transform all at once, it is easier to express the transformation
$\phi^{*}$ as a composition of $L$ smooth, one-to-one transformations, such that:
$ \phi^{*} = \phi_{L} \circ \phi_{L-1} \circ \dots \circ \phi_{1}.$ Each component transformation $\phi_{l}$ is chosen to nudge the distribution closer to the target $\rho^{*}$. Formally, each transformation is chosen as a slight perturbation of the identity transform in andescent direction of the KL-divergence chosen to approximate its gradient in a specified metric space over distribution. Formally, each component step is an descent step if:
$$
\text{KL}(\rho_{\phi^{l+1}} \parallel \rho^{*}) \leq \text{KL}(\rho_{\phi^{l}} \parallel \rho^{*}),
$$
where $\phi^{l} = \phi_{l} \circ \dots \circ \phi_{1}$ is the accumulation of first $l$ maps.

If we generalize the discrete index $l$ to continuous artificial time $t$, then the distribution  evolves continuously in $t$. For simplicity, we denote this evolving distribution as $\rho_{t}$. The decreasing KL-divergence implies that its derivative with respect to $t$ is non-positive, i.e
$$
\frac{d}{dt} \text{KL}(\rho_{t} \parallel \rho^{*}) \leq 0.
$$
The one-to-one smooth transformation can be accomplished by solving a ODE, such that 
$$
\frac{d\phi(\cdot, t)}{dt} = u(\phi(\cdot, t),t),
$$
for some velocity field $u$. 

Thus, the problem of finding the optimal distribution approximation with respect to the KL-divergence reduces to determining the optimal velocity field. Specifically, the steepest descent direction for minimizing $\text{KL}(\rho_{\phi} \parallel \rho^{*})$ corresponds to the variation of the KL divergence with respect to the smooth transformation $\phi$. This variation provides a principled way to construct the velocity field that drives the distribution  $\rho_t$ closer to $\rho^{*}$ over time. %By leveraging this steepest descent direction, we can iteratively update the velocity field to ensure a systematic reduction in the KL divergence, leading to the desired approximation of the target distribution. 

As shown in \cite{wang2021particle}, the variation of the KL-divergence with respect to the flow map $\phi$ is:
\begin{equation}\label{kl}
-\frac{\partial D_{KL}(\rho_{\phi} \parallel \rho^{*})}{\partial \phi}
= - \nabla \rho_{\phi}(x) + \rho_{\phi}(x) \nabla \log \rho^{*}(x).
\end{equation}
In practice, we only have $N$ samples $\{x_{i}\}_{i=1}^{N}$ instead of the variational distribution $\rho(x)$. So we are forced to estimate the variational density from the empirical distribution of its samples, $\hat{\rho} = \frac{1}{N}\sum_{i=1}^{N} \delta_{x_{i}}$. This approach is problematic since$\nabla\hat{\rho}$ is not well-defined since the indicator function $\delta$ is not differentiable. To apply Equation \eqref{kl}, convolve the right-hand side by a kernel function $\kappa$, after approximating $\rho$ by $\hat{\rho}$. This leads to the formulation of Stein Variational Gradient Descent (SVGD) \cite{liu2016stein,wang2021particle}, where the vector field driving $\phi$ is:
\begin{equation}\label{SVGD}
\phi^{*}(x) = -\sum_{j=1}^{N} \nabla_{x_j} \kappa(x_j, x) - \sum_{j=1}^{N}  \kappa(x_j, x) \log \rho^{*}(x_j).
\end{equation}
Equation \eqref{SVGD} is, equivalently, a sample approximation to the gradient of the KL divergence in the reproducing norm associated with the reproducing kernel Hilbert space built from the kernel $\kappa$ \cite{liu2016stein}.

Then, given a learning rate $\eta$, SVGD updates each sample with a simple forward Euler step:
\begin{equation} \label{eqn: SVGD Euler}
    x_{i} \leftarrow x_{i} - \eta \phi^{*}(x_{i}).
\end{equation}

Equations \eqref{SVGD} and \eqref{eqn: SVGD Euler} define an intuitive particle dynamic. The second term guides particles toward regions of higher density in the target distribution $\rho^{*}$. Intuitively, particles are ``attracted" to areas where 
$\rho^{*}(x)$ is large, aligning the particle distribution with the target density. The first term introduces a repulsive force between particles that depends on the kernel $\kappa$. This term prevents the particles from clustering too closely. The repulsion is stronger when particles are closer together, as dictated by the gradient of the kernel. Without the repulsive interaction term, all of the particles would converge to the local maximizers of the target. By balancing repulsive interactions between particles, and each particle's desire to maximize density, SVGD promotes particle distributions that mirror the underlying target distribution.

%\red{Gaussian Kernel}

In practice, a Gaussian kernel, is typically used:
\begin{equation}
\kappa(x,x^{\prime};h) = \frac{1}{\sqrt{(2 \pi)^d} h^d} \exp\left( -\frac{1}{2} \frac{\|x - x^{\prime}\|_2^2}{h^2} \right)
\end{equation}
where $h$ is the bandwidth of Gaussian kernel and $\|\cdot\|_2$ denotes Euclidean norm.

%\red{Median Trick for bandwidth}

The bandwidth parameter, $h$, is crucial for the algorithm’s efficiency and performance since it is responsible for establishing the balance between the repulsive interactions and the density maximizing forces. It directly impacts both the quality of the approximation to the target distribution and the convergence behavior. Selecting an appropriate bandwidth, however, can be challenging without sufficient prior knowledge of the target distribution.

Two widely used strategies for setting the bandwidth have emerged in the literature. In the first approach, the bandwidth remains constant throughout the optimization process \cite{wang2021particle}. While simple, this method requires careful tuning as a fixed bandwidth may not adapt well to the dynamic nature of particle interactions during SVGD. The choice of this constant can be problem-dependent and may require cross-validation or other tuning methods. 

A more adaptive and commonly employed method is the median trick, where the bandwidth 
$h$ is updated dynamically based on the particle distribution. Specifically, the bandwidth is set as:
$$
h = \text{med}^{2} / \log N.
$$
where $\text{med}$ denotes the median of the pairwise distances between particles $\{x_{i}\}_{i=1}^{N}$ \cite{liu2016stein}. 

The median trick is a heuristic and does not come with formal theoretical guarantees regarding convergence rates or the quality of the approximation. We will illustrate two simple cases where this approach fails. First, in a multimodal distribution with low variance within each mode but large distances between the modes, the median trick overestimates the bandwidth due to the separation between modes, causing poor resolution of individual modes, see table \ref{Gaussian Mixture with Unequal Weights} and figure \ref{fig:2D2modes}. Second, in a unimodal bivariate Gaussian distribution with significantly different scales of standard deviations along the two axes, the median trick underestimates the bandwidth for one axis and overestimates it for the other, as shown in table \ref{tb:AnisotropicGaussian}. This mismatch results in a poor approximation of the target distribution. Since many real posteriors are both highly eccentric, and multimodal, adopting a heuristic based on the distribution of distances between samples is unreliable.

\subsection{Bandwidth Update}
\label{sec:bandwidth}
Given the importance of selecting an appropriate bandwidth, we introduce a gradient-based method to adaptively update the bandwidth for optimal alignment with the target distribution.

The objective is to determine the bandwidth $h$ that minimizes the KL-divergence between the variational density and the target distribution. The variational density, convolved with a Gaussian kernel, is defined as: 
$$
\rho_{h}(x) = \frac{1}{N}\sum_{i=1}^{N} \frac{1}{\sqrt{(2 \pi)^n} h^n} \exp\left( -\frac{1}{2} \frac{\|x - x_i\|_2^2}{h^2} \right).
$$
where $\rho_{h}$ depends on the bandwidth $h$. This density serves as the empirical distribution smoothed by the kernel. Consequently, the KL-divergence becomes a function of $h$, and the goal is to optimize 
$h$ such that:
$$
h^{*} = \argmin_{h} \{\text{KL}(\rho_{h} \parallel \rho^{*})\}.
$$

The gradient of the KL divergence with respect to $h$ can be expressed as:
\begin{equation}\label{bd_gradient}
    \frac{\partial D_{KL}(\rho_{h} \parallel \rho^{*})}{\partial h} 
    =
    \frac{1}{h} \int_x \log\left({\frac{\rho_{h}(x)}{\rho^{*}(x)}} \right)  \frac{1}{N} \sum_{i=1}^{N} \left( \frac{\|x-x_{i}\|_{2}^{2}}{h^{2}} - n \right) \kappa(x,x_i;h) dx
\end{equation}
where $\kappa$ is the Gaussian kernel function.
For details, please refer to section \ref{appx:bd} in the Appendix.

The integral in Equation \eqref{bd_gradient} can be expressed as an expectation against the variational density by dividing and multiplying by $\rho_{h}$:
\begin{equation}\label{eq:bandwidth_update_unsimplified}
\frac{\partial D_{KL}(\rho_{h} \parallel \rho^{*})}{\partial h} = \frac{1}{h}  \mathop{\mathbb{E}_{X \sim \rho_{h}} }\left[ \frac{1}{\rho_{h}(X)} \log\left({\frac{\rho_{h}(X)}{\rho^{*}(X)}}\right) \frac{1}{N} \sum_{i=1}^{N} \left( \frac{\|X-x_{i}\|_{2}^{2}}{h^{2}} - n \right) \kappa(X,x_i;h) \right].
\end{equation}

To approximate the expectation in Equation \eqref{eq:bandwidth_update_unsimplified}, we can use the sample average evaluated at the set of samples drawn from the distribution $\rho_{h}$. The remaining task is to generate such samples efficiently.

Since $\kappa$ is chosen as a Gaussian kernel, the variational density $\rho_{h}$ is a Gaussian mixture model with each Gaussian component centered at $x_i$ and assigned equal weight. As a result, to generate samples from $\rho_{h}$, we can use a two-step sampling process: 
\begin{enumerate}
    \item Sample an index $J$ uniformly from $\{1, \dots, N\}$, to select a Gaussian component.
    \item Conditionally on the chosen index $J$, sample $X$ from the Gaussian distribution with mean $x_{J}$ and covariance matrix $h^{2}I$:
    $$
    X\mid J \sim \mathcal{N}(x_J, h^{2}I).
    $$
\end{enumerate}

To interpret Equation \eqref{eq:bandwidth_update_unsimplified}, we expand it according to the two-stage sampling process outlined above. Expanding the expectation gives:
\begin{equation}
\begin{aligned}
& \frac{\partial D_{KL}(\rho_{h} \parallel \rho^{*})}{\partial h}  \\
& = \frac{1}{h} 
\mathop{\mathbb{E}_{J \sim \text{Unif}[1,N]}} \left[\mathop{\mathbb{E}_{X \sim \mathcal{N}(x_J, hI)}}\left[ \log\left({\frac{\rho_{h}(X)}{\rho^{*}(X)}}\right) \frac{1}{N} \sum_{i=1}^{N} \left( \frac{\|X-x_{i}\|_{2}^{2}}{h^{2}} - n \right) \frac{\kappa(X,x_i;h)}{\rho_{h}(X)} \right] \right].
\end{aligned}
\end{equation}
The first term inside the average is the log-ratio of the kernel and the target density. Equivalently, this equals the negative logarithm of the importance weight had we used an importance weighting scheme to approximate the target given samples from the variational density. This ratio is positive where the variational density overestimates the target, and negative where it underestimates the target.
The second term measures whether the distance between $X$ and $x_i$ is larger, or smaller than expected under a sample draw from a Gaussian with bandwidth $h$. The ratio at the end is the conditional probability that a sample at location $X$ was drawn from a mode centered at $x_i$ given the positions of all the particles. 

There are four cases here: over/under estimate the target, and whether the set of particles are more/less dense about $x$ than is suggested by the bandwidth, corresponding to the sign of the first term and the second term in the expectation.
\begin{enumerate}
    \item Over estimate and closer than the bandwidth implies that the estimate density is too high near the mode, suggesting that the estimate density is too concentrated and bandwidth $h$ should be increased.
    \item Under estimate and further than the bandwidth implies that the estimate density is too low in the tails, suggesting that the estimate density is too concentrated  and bandwidth $h$ should be increased.
    \item Over estimate and further than the bandwidth implies that the estimate density is too high in the tails, suggesting that the estimate density is too diffuse and bandwidth $h$ should be decreased.
    \item Under estimate and closer than the bandwidth implies that the estimate density is too low near the mode, suggesting that the estimate density is too diffuse and bandwidth $h$ should be increased.
\end{enumerate}
Then, we take the average over $X$.

%\red{Back to hierarchical model, where we can define two bandwidth}

So far, we have used a spherically symmetric kernel with covariance matrix of the form $\Sigma = h^{2}I_{n}$. The hierarchical model defines a joint posterior over a pair of distinct variables, $(x,\theta) \in \mathbb{R}^{2n}$. As $x$ and $\theta$ play different roles, they may vary over different scales. So, in the hierarchical setting, we use separate bandwidths for $x$ and $\theta$. We adopt a covariance matrix of the form:
$$
\Sigma = \begin{pmatrix}
    h_{1}^{2} I_{d} & 0 \\
    0 & h_{2}^{2} I_{d}
\end{pmatrix}.
$$
%%Then the Gaussian distribution centered at $(x_{i}, \theta_{i})$ has density function 
%$$
%\frac{1}{(2\pi)^{d}h_{1}^{d}h_{2}^{d}}\exp(-\frac{\|x-x_i\|_{2}^{2}}{2h_{1}^{2}} - \frac{\|\theta - \theta_{i}\|_{2}^{2}}{2h_{2}^{2}}).
%$$

In this case:
$$
\rho = \frac{1}{N} \sum_{i=1}^{N} \frac{1}{(2\pi)^{d}h_{1}^{d}h_{2}^{d}}\exp\left(-\frac{\|x-x_i\|_{2}^{2}}{2h_{1}^{2}} - \frac{\|\theta - \theta_{i}\|_{2}^{2}}{2h_{2}^{2}} \right),
$$
and,
\begin{equation}
\begin{aligned}
& \frac{\partial D_{KL}(\rho \parallel \rho^{*})}{\partial h_1} \\ 
& = \int \ln \frac{\rho(x)}{\rho^*(x)} \frac{1}{N} \sum_{i=1}^{N} \frac{1}{h_1}\left(\frac{\|x-x_{i}\|_{2}^{2}}{h_{1}^{2}} - n \right) \frac{1}{(2\pi)^{n}h_{1}^{n}h_{2}^{n}}\exp\left(-\frac{\|x-x_i\|_{2}^{2}}{2h_{1}^{2}} - \frac{\|\theta - \theta_{i}\|_{2}^{2}}{2h_{2}^{2}} \right), \\
&  \frac{\partial D_{KL}(\rho \parallel \rho^{*})}{\partial h_2}\\  
& = \int \ln \frac{\rho(x)}{\rho^*(x)} \frac{1}{N} \sum_{i=1}^{N} \frac{1}{h_2}\left(\frac{\|\theta-\theta_{i}\|_{2}^{2}}{h_{2}^{2}} - n \right) \frac{1}{(2\pi)^{n}h_{1}^{n}h_{2}^{n}}\exp\left(-\frac{\|x-x_i\|_{2}^{2}}{2h_{1}^{2}} - \frac{\|\theta - \theta_{i}\|_{2}^{2}}{2h_{2}^{2}} \right).
\end{aligned}
\end{equation}

We adopt the same conditional sampling approach to approximate each integral.

%\red{Alternative update $\log h$}

When bandwidth $h$ is small, direct updates to $h$ can risk overstepping and producing negative values. To avoid constrained optimization, we update $\log h$ instead. Then Equation \eqref{bd_gradient} is replaced with:
\begin{equation}
\frac{\partial D_{KL}(\rho \parallel \rho^{*})}{\partial \log h} 
=
\int \ln{\frac{\rho(x)}{\rho^{*}(x)}} \frac{1}{N} \sum_{i=1}^{N} \left( \frac{\|x-x_{i}\|_{2}^{2}}{h^{2}} - n \right) \kappa(x,x_i;h) dx.
\end{equation}

By removing the factor $\frac{1}{h}$, this approach mitigates the risk of overstepping for small $h$. %Additionally, since $\log h$ is unconstrained, it can be robustly updated using gradient-based methods without requiring additional constraints.

\subsection{Birth-Death Process}
\label{sec:birthdeath}
A Birth-Death process is a  Markov process used to model systems in which the number of entities, such as individuals in a population or particles in a system, evolves through two types of events: births, in which an individual is added to the population, and deaths, in which an individual is removed. These processes are characterized by random transitions between states, with birth and death rates governing the probabilistic occurrence of these events.
Birth-Death Process have been used to accelerate sampling, especially in Langevin Dynamics \cite{lu2019acceleratinglangevinsamplingbirthdeath}.

Birth-Death processes can rapidly move probability mass directly between modes by removing particles from oversampled modes and adding them to undersampled modes. This effectively jumps, or teleports, particles between modes. Since the particles do not have to move continuously they can easily skip over low-probability regions. In contrast most MCMC methods, or EVI methods, that use local updates, struggle to exchange mass between separated modes. 

Following \cite{lu2019acceleratinglangevinsamplingbirthdeath}, we consider a Birth-Death equation 
\begin{equation}
\partial_{t} \rho_{t}(x) = -\alpha_{t}(x)\rho_{t}(x)\end{equation}
where the $\alpha_t(x)$ is the difference in the per capita death and birth rates for individuals of type $x$. The difference is set to: 
$$\alpha_{t}(x) = \log\left(\frac{\rho_t(x)}{\rho^*(x)} \right) - \int \log\left(\frac{\rho_t(x)}{\rho^*(x)} \right) \rho_t dx.
$$

The first term ensures that the per capita growth rate due to births and deaths at $x$ is negative if the variational density greatly overestimates the target at $x$. If, instead, the variational density greatly underestimates the target, then the per capita growth rate is positive. The second term conserves the total mass of the density $\rho_t$. As before, the integral may be expressed as an expectation against the variational density, then approximated using Monte Carlo sampling from the variational density.

Due to the limitations imposed by finite samples in practice, a particle-based birth-death process is required. Similar to SVGD, we convolve the empirical process with a smooth kernel $K$, specifically the Gaussian kernel $\kappa$ in implementation. In other words, we use the same variational distribution $\rho_{h}(x)$ as in SVGD. After applying this transformation, the explicit formula for $\alpha_{t}(x)$ becomes:
$$
\alpha(x) = \log \left(\frac{\rho_{h}(x)}{\rho^{*}(x)} \right) - \frac{1}{N} \sum_{i=1}^{N} \log \left(\frac{\rho_{h}(x_{i})}{\rho^{*}(x_{i})} \right).
$$
Then given a time step $\Delta t$, each sample $x_i$ has probability 
$$
1-\exp(-\hat{\alpha}(x_i)\Delta t)
$$
to be killed, if $\hat{\alpha}(x_i) > 0$. Otherwise, $x_i$ has probability
$$
1-\exp(\hat{\alpha}(x_i)\Delta t)
$$
to be duplicated. Here $\hat{\alpha}(x_{i}) = \alpha(x_{i}) - \frac{1}{N} \sum_{i=1}^{N} \alpha(x_{i})$, represents the deviation of $\alpha(x_{i})$ from its mean.

%\section{Methods}
%In this section, we will introduce two key algorithms: Birth-Death sampling and Backtracking, both of are employed to improve sampling efficiency and accuracy in high-dimensional settings. These methods play a critical role in enhancing exploration of the target distribution by re-weighting between different modes and enforcing robust updates on samples. We summarize our complete method at the end of this section.

%\subsection{Birth-Death Sampling}
\begin{algorithm}[h]
\caption{Birth-Death Sampling}
\label{alg:bd}
\begin{algorithmic}
\FOR{$i \leftarrow 1$ to $N$}
    \STATE $\alpha_i = \log \rho(x_i) - \log \rho^{*}(x_i)$
\ENDFOR
\STATE $\bar{\alpha} = \frac{1}{N}\sum_{i=1}^{N}\alpha_{i}$
\FOR{$i \leftarrow 1$ to $N$}
    \IF{$\alpha_{i} > \bar{\alpha}$}
        \STATE $s \sim \mathcal{B}(1-\exp((\bar{\alpha} - \alpha_{i})\Delta t))$
        \IF{$s == 1$}
            \STATE sample $\sigma \sim \mathcal{N}(0, h^{2}I_{d})$
            \STATE $J$ uniformly chosen from $\{1, \dots, i-1, i+1, \dots, N\}$
            \STATE $x_i \leftarrow x_J + \sigma$
        \ENDIF
        % \STATE kill $x_i$ with probability $1-\exp((\bar{\alpha} - \alpha_{i})\Delta t)$
        % \STATE sample $\sigma \sim \mathcal{N}(0, hI_{d})$
        % \STATE generate one sample $x_J + \sigma$ with $J$ uniformly chose from $\{1, \dots, i-1, i+1, \dots, N\}$
    \ELSE
         \STATE $s \sim \mathcal{B}(1-\exp((\alpha_{i} - \bar{\alpha})\Delta t))$
        \IF{$s == 1$}
            \STATE sample $\sigma \sim \mathcal{N}(0, h^{2}I_{d})$
            \STATE $J$ uniformly chosen from $\{1, \dots, i-1, i+1, \dots, N\}$
            \STATE $x_J \leftarrow x_i + \sigma$
        \ENDIF
        % \STATE duplicate $x_i$ with probability $1-\exp((\alpha_{i} - \bar{\alpha})\Delta t)$
        % \STATE kill one
    \ENDIF
\ENDFOR
\end{algorithmic}
\end{algorithm}

To preserve the total population size during the Birth-Death process, we uniformly resample a new particle any time one is killed, and uniformly sample a particle to remove if one is added. To ensure that new particles are not placed on top of existing particles, we resample from the variational distribution. Since variational empirical distribution is approximated as Gaussian mixture with a mode located at each sample, we resample by selecting a parent particle from the current population, then add a Gaussian noise with mean $0$ and variance $h^{2}I_{d}$. The proposed approach is outlined in Algorithm \ref{alg:bd}, where $\mathcal{B}$ denote Bernoulli distribution.

\subsection{Back Tracking}
\label{sec:backtrack}
To prevent over-stepping during the update process, we append a Back-Tracking procedure to each proposed update (SVGD, bandwidth, Birth-Death). First, the KL divergence is approximated after each update. If the KL divergence decreased, the update is accepted, and the original value is replaced. If not, the step size is halved, and the update is re-evaluated. This process is repeated until the update results in a non-increasing KL divergence. As the exact KL divergence is typically intractable, it is approximated empirically by substituting the expectation with a sample mean over many draws from the variational distribution.  Note, this Back-Tracking procedure is not exact, since it uses approximate, noisy, KL evaluations.

Algorithm \ref{alg:bt} explains the Back-Tracking procedure. Back-tracking is applied for the bandwidth update using the same procedure. 

\begin{algorithm}[h]
\caption{Sample KL divergence}
\label{alg:kl}
\begin{algorithmic}
\STATE \textbf{Input:} Samples  $\{x_{i}\}_{i=1}^{N}$, Kernel $K$, target distribution $\rho_{*}$
\STATE
\textbf{Return:}
$k = \frac{1}{N}\sum_{i=1}^N \log(\frac{1}{N}\sum_{j=1}^{N}K(x_i, x_j)) - \log(\rho^{*}(x_i))$
\end{algorithmic}
\end{algorithm}

\begin{algorithm}[h]
\caption{Sample Update with Back-Tracking}
\label{alg:bt}
\begin{algorithmic}
\STATE \textbf{Input:}  Original samples $\{x_{i}\}_{i=1}^{N}$, updated samples $\{\tilde{x_{i}}\}_{i=1}^{N}$
\STATE $k = \frac{1}{N}\sum_{i=1}^N \log(\frac{1}{N}\sum_{j=1}^{N}K(x_i, x_j)) - \log(\rho^{*}(x_i))$
\STATE $\tilde{k} = \frac{1}{N}\sum_{i=1}^N \log(\frac{1}{N}\sum_{j=1}^{N}K(\tilde{x_i}, \tilde{x_j})) - \log(\rho^{*}(\tilde{x_i}))$
\WHILE{$\tilde{k} > k$}
    \FOR{$i \leftarrow 1$ to $N$}
        \STATE $\tilde{x_{i}} \leftarrow x_i + (\tilde{x_i} - x_i) / 2$
    \ENDFOR
    \STATE $\tilde{k} = \frac{1}{N}\sum_{i=1}^N \log(\frac{1}{N}\sum_{j=1}^{N}K(\tilde{x_i}, \tilde{x_j})) - \log(\rho^{*}(\tilde{x_i}))$
\ENDWHILE
\FOR{$i \leftarrow 1$ to $N$}
    \STATE $x_i \leftarrow \tilde{x_i}$
\ENDFOR
\end{algorithmic}
\end{algorithm}

To append Back-Tracking to the Birth-Death process we use a two-step procedure. We evaluate the KL-divergence at two checkpoints. The first occurs when a sample is removed, and a parent is uniformly selected from the remaining particles or a sample is generated with one uniformly sample removed. The second occurs when Gaussian noise is added to move the child away from their parent. If the KL-divergence increases at the first checkpoint, we revert to the original sample. At the second checkpoint, if the KL-divergence increases, we halve the added noise. This process is repeated until the estimated KL decreases.

\begin{algorithm}[h]
\caption{Birth-Death with Back-Tracking}
\label{alg:bd+bt}
\begin{algorithmic}
\STATE
\textbf{Inputs:}
Samples $\{x_{i}\}_{i=1}^{N}$, sample to be killed $x_{j}$, sample to be generated $x_{j^\prime}$
\STATE
$k = \text{Sample KL divergence}(x_{1},\dots, x_{j-1}, x_{j}, x_{j+1}, \dots, x_{N})$ 
\STATE
$\tilde{k} = \text{Sample KL divergence}(x_{1},\dots, x_{j-1}, x_{j^\prime}, x_{j+1}, \dots, x_{N})$ 
\IF{$\tilde{k} < k$}
    \STATE $x_{j} \leftarrow x_{j^\prime}$
    \STATE sample $\sigma \sim \mathcal{N}(0, h^{2}I_{d})$
    \STATE sample $\sigma \sim \mathcal{N}(0, h^{2}I_{d})$
    \STATE $x^\prime \leftarrow x_{j} + \sigma$
    \STATE $k = \text{Sample KL divergence}(x_{1},\dots, x_{j-1}, x_{j}, x_{j+1}, \dots, x_{N})$ 
    \STATE $\tilde{k} = \text{Sample KL divergence}(x_{1},\dots, x_{j-1}, x^\prime, x_{j+1}, \dots, x_{N})$ 
    \WHILE{$\tilde{k} > k$}
        \STATE $\sigma \leftarrow \sigma / 2$
        \STATE $x^\prime \leftarrow x_{j} + \sigma$
        \STATE $\tilde{k} = \text{Sample KL divergence}(x_{1},\dots, x_{j-1}, x^\prime, x_{j+1}, \dots, x_{N})$ 
    \ENDWHILE
    \STATE $x_{j} \leftarrow x^{\prime}$
\ENDIF

\end{algorithmic}
\end{algorithm}

\section{Experiments}
\label{sec:experiment}

\subsection{2D Gaussian}

\subsubsection{Gaussian Mixture with Unequal Weights}
In this section, we consider an example of 2D Gaussian mixture distribution with unequal weights. We set our target distribution to be
$\boldsymbol{y}_i \sim \frac{1}{4} \mathcal{N}\left([-4,-4]^{\top}, I\right)+ \frac{3}{4} \mathcal{N}\left([4,4]^{\top}, I\right)$. We use N = 100 particles and the initial particles are sampled from two-dimensional standard Gaussian distribution. 

\begin{figure}[htbp]
    \centering
    
    % First Row
    \begin{subfigure}[b]{0.45\textwidth}
        \centering
        \includegraphics[width=\textwidth]{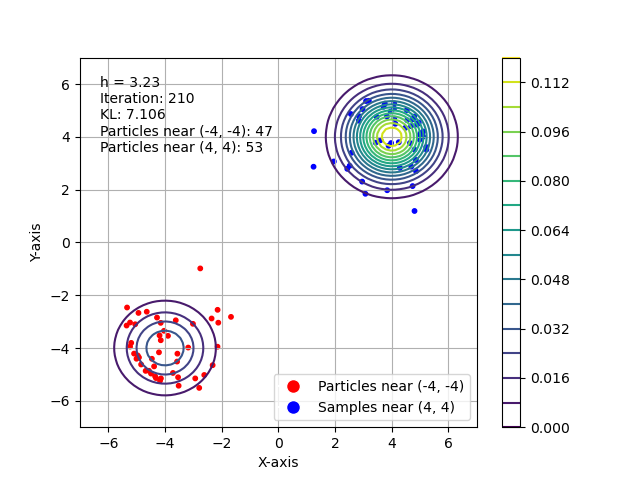}
        \caption{SVGD with an adaptive bandwidth using median trick}
        \label{fig:sub1}
    \end{subfigure}
    \hfill
    \begin{subfigure}[b]{0.45\textwidth}
        \centering
        \includegraphics[width=\textwidth]{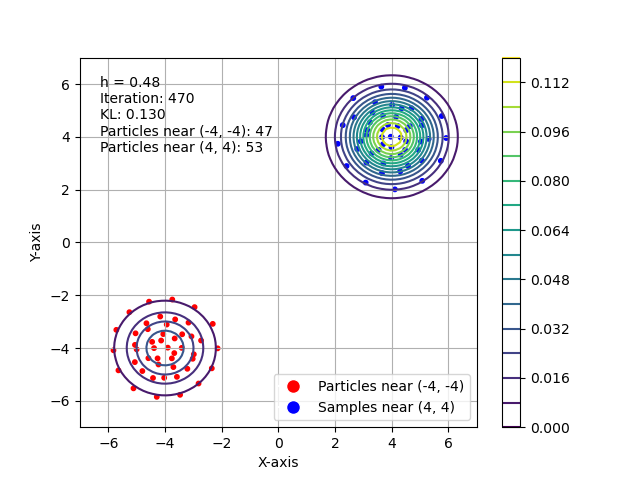}
        \caption{SVGD with an adaptive bandwidth using gradient descent}
        \label{fig:sub2}
    \end{subfigure}
    
    \vspace{1em} % Space between rows
    
    % Second Row
    \begin{subfigure}[b]{0.45\textwidth}
        \centering
        \includegraphics[width=\textwidth]{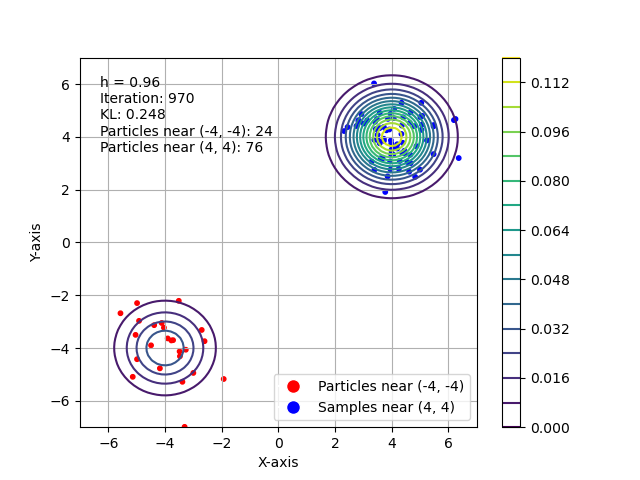}
        \caption{Birth-death with an adaptive bandwidth using median trick}
        \label{fig:sub3}
    \end{subfigure}
    \hfill
    \begin{subfigure}[b]{0.45\textwidth}
        \centering
        \includegraphics[width=\textwidth]{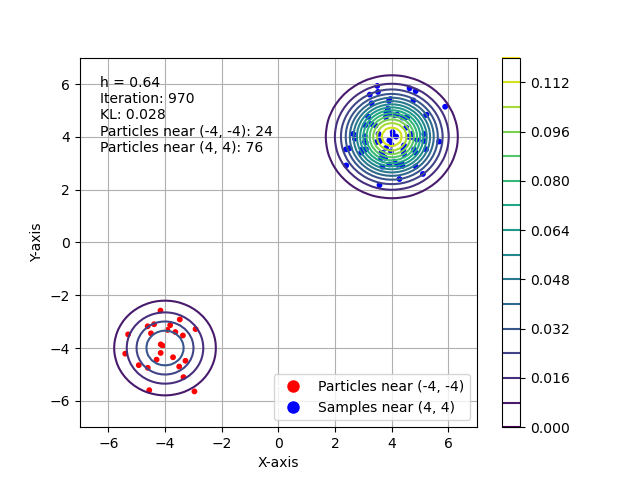}
        \caption{Birth-death with an adaptive bandwidth using gradient descent}
        \label{fig:sub4}
    \end{subfigure}
    
    \vspace{1em} % Space between rows
    
    % Third Row
    \begin{subfigure}[b]{0.45\textwidth}
        \centering
        \includegraphics[width=\textwidth]{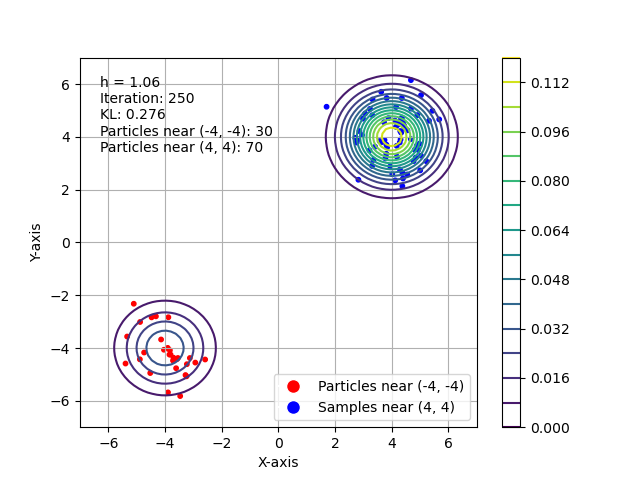}
        \caption{SVGD+Birth-death with an adaptive bandwidth using median trick}
        \label{fig:sub5}
    \end{subfigure}
    \hfill
    \begin{subfigure}[b]{0.45\textwidth}
        \centering
        \includegraphics[width=\textwidth]{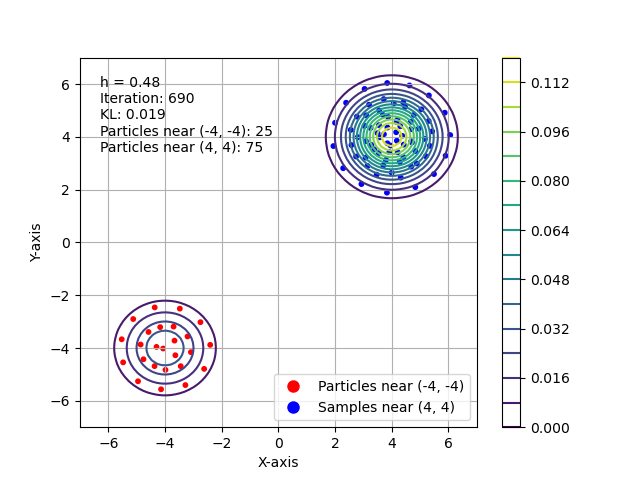}
        \caption{SVGD+Birth-death with an adaptive bandwidth using gradient descent}
        \label{fig:sub6}
    \end{subfigure}
    
    \caption{Particles obtained by SVGD (first row), Birth-death (second row) and SVGD+Birth-death (third row) with an adaptive bandwidth using the median trick (left column) and the gradient descent (right column).     }
    \label{fig:2D2modes}
\end{figure}

The primary challenge in this example is transferring weights between two well-separated modes modes. The sampling results returned by different methods are
shown in Figure \ref{fig:2D2modes}. We observe that SVGD struggles to transfer weight between the modes, as it tends to assign particles only to nearby modes, thereby failing to allocate the appropriate number of particles in each mode. The guide particles largely collect in the modes based on their initial position, so the resulting estimate is strongly initialization dependent. For example, Figure \ref{fig:2D2modes} shows that, both SVGD, and SVGD with an adaptive bandwidth, assign about half of the particles to each mode when the particles are initialized from a distribution that is symmetric about the origin, and places roughly equal mass in the basin of attraction of each mode (see panels (a) and (b)). In contrast, when Birth-Death is introduced, 24 to 30\% particles are assigned to the mode that represented 25\% of the probability mass, while the remaining 70 to 76\% converge to the mode containing 75\% of target mass.

In contrast, the birth-death method directly adjusts the number of samples in each mode by duplicating or eliminating samples, enabling a faster initial decrease in KL divergence. We also note that SVGD is more effective at locally approximating the distribution due to its access to gradient information and deterministic optimization procedure. These result in regular, lattice-like allocation of particles that produce smooth, reasonably symmetric, variational distributions. 

The KL-based gradient descent bandwidth update consistently outperforms the median trick heuristic for bandwidth selection. It results in a lower KL divergence and exhibits greater robustness to initialization. Our experiments also indicates that the median trick heuristic is prone to divergence when subjected to poor initialization. In this example, the median trick overestimates the necessary bandwidth, since the median distance between particles is largely driven by the distance between the modes, rather than the standard deviation of each mode. The kernel bandwidth should be adapted to match the standard deviation within each mode. In the example shown, methods using the median trick returned bandwidths randing from 0.96 to 3.23. In particular, SVGD with the median trick returns $h = 3.23$. This leads to systematic errors in the estimated shapes of each mode, since the particles are pushed away from the other mode, and concentrate on the side of each mode farthest from the other mode (see Figure \ref{fig:2D2modes} panel (a)). In contrast, when the bandwidth is optimized for directly, it returned values between 0.48 and 0.62. 

\begin{table}[h]
\resizebox{\columnwidth}{!}{
\begin{tabular}{lllll}
\hline
\centering
                & KL & KL(10)    & \# iterations   & running time   \\ \hline\hhline{=====}
h(m) + SVGD      & 1.4 (1.2, 1.3)  & 12 (11.4, 12.5) & 591 (382, 812)                            & 26.6 (10.8, 49.7)         \\ \hline
h(g) + SVGD    & 0.27 (0.24, 0.30)  & 5.1 (3.5, 5.7) & 220 (131, 352)                            & 19.0 (10.4, 31.3)         \\ \hline\hhline{=====}

h(m) + BD         & diverge & diverge & diverge                     & diverge       \\ \hline
h(g) + BD         & \textbf{0.025 (0.006, 0.073)}  & 0.12 (0.031, 0.28) & 119 (22, 262)                  & \textbf{13.3 (6.5, 25.3)}      \\  \hline
h(g) + BD(b)      & 0.028 (0.006, 0.089)   & 0.14 (0.042, 0.34) & 132 (32, 292)                             & 18.9 (9.2, 34.9)          \\ \hline\hhline{=====}

h(m) + SVGD + BD       & diverge   & diverge  & diverge                              & diverge          \\ \hline
h(g) + SVGD + BD   & 0.026 (0.009, 0.045)  & \textbf{0.080 (0.024, 0.224)} & \textbf{113 (22, 252)}                            & 15.1 (8.0, 26)          \\ \hline
h(g) + SVGD + BD(b) & \textbf{0.022 (0.012, 0.038)}  & 0.099 (0.033, 0.24) & \textbf{112 (22, 260)}            
                & \textbf{12.4 (6.0, 19.3)}        \\ \hline

\end{tabular}}
\caption{Gaussian Mixture with Unequal Weights. 
\textit{Rows:} SVGD, Birth-Death (BD), and Mixed Method. h(m) denotes bandwidth update using the median trick, while h(g) uses gradient descent. BD(b) indicates backtracking in Birth-Death. \textit{Columns:} KL divergence, KL after 10 iterations, total number of iterations, and running time.  Each numerical entry shows the mean followed by a 95\% confidence interval in parentheses. Bolded values represent the best or near-best entries in each column.}

\label{Gaussian Mixture with Unequal Weights}
\end{table}

 The convergence rates for each method are summarized in Table \ref{Gaussian Mixture with Unequal Weights}. Note that the Birth-Death method achieves significantly lower final KL divergence, and converges faster than the SVGD method due to its efficient mass transfer between the two modes. The mixed method further reduces KL divergence slightly compared to the Birth-Death method by leveraging the gradient-based optimization properties of SVGD to refine its approximations to the shape of each mode. It runs in essentially the same time as the pure Birth-Death process, as measured in iteration count, and wall time.

To conclude, the mixed method, combines the strengths of both approaches, ensuring accurate weight allocation for each mode, efficient mass exchange between modes, accurate bandwidth estimation, and locally precise approximations. Notice that, no method using the median trick to select the bandwidth succeeds. 

\subsubsection{Anisotropic Gaussian}

In this section, we analyze an anisotropic Gaussian distribution with a significantly different variances along its principal axes. We set our target distribution to be $
\boldsymbol{y}_i \sim  \mathcal{N} \left(
\begin{bmatrix} 0 \\ 0 \end{bmatrix}, 
\begin{bmatrix} 0.1 & 0 \\ 0 & 1 \end{bmatrix}
\right) $. We use $N = 100$ particles and the initial particles are sampled from the two-dimensional standard Gaussian distribution.

The challenge in this example lies in accurately capturing the elongated and narrow structure of the distribution. This example is chosen to illustrate the importance of selecting the bandwidth correctly. The results are shown in Table \ref{tb:AnisotropicGaussian}.

 Note that updating the bandwidth enhances the performance of both methods. The median trick returns exceptionally poor estimates, since it overestimates the necessary bandwidth. It selects a bandwidth on the scale of the larger variance. This is inappropriate since the direction of larger variance can be captured by distributing particles along that direction, each with a small bandwidth that captures the spread along the minor axis. In other words, the median trick chooses the wrong scale heuristic. It focuses on an intermediate scale, rather than the standard deviation along the minor axis. As a result, it necessarily over-covers the target. Optimizing the bandwidth resolves this problem.
 
 With bandwidth updating, the SVGD method outperforms the Birth-Death method. This aligns with the fact that SVGD is an gradient-based method, allowing it to achieve a more accurate representation of the target distribution, particularly when the distribution has a non-uniform shape. It's not surprising that the SVGD method achieves a KL divergence close to that of the mixed method, as the distribution is unimodal, so there is no need to teleport mass between separate regions. However, the number of iterations and running time indicate that the Birth-Death method still accelerates convergence by allowing particles to jump towards regions of high density when the variational distribution is a poor cover of the target (e.g. at initialization). In all cases, adding a backtracking step improved the final KL estimate.

\begin{table}[h!]
\centering
\resizebox{\columnwidth}{!}{
\begin{tabular}{llllll}
\hline
                & KL & KL(10)    & \# iterations   & running time   \\ \hline\hhline{======}
 h(m) + SVGD      & 26 (15,35)  & 61 (57,64) & 404 (112,852)                            & 7.5 (3.0,15.1)         \\ \hline
h(g) + SVGD      & \textbf{0.049 (0.036,0.064)}  & 5.1 (2.1,7.3) & 270 (192,352)                            & 6.5 (4.6,8.8)         \\ \hline\hhline{======}

h(m) + BD         & diverge & diverge&  diverge                     & diverge       \\ \hline
h(g) + BD         & 0.19 (0.14,0.25)  &3.2 (0.88,5.6)& \textbf{159 (72,282)}                 &    \textbf{4.0 (2.5, 6.2)}      \\  \hline
h(g) + BD(b)      &  0.15 (0.092,0.24)   & \textbf{2.2 (0.71,4.6)} & 159 (62,312)                            &  5.5 (3.2,9.5)          \\ \hline\hhline{======}

h(m) + SVGD + BD       & diverge   & diverge  & diverge                              & diverge          \\ \hline
h(g) + SVGD + BD   & 0.064 (0.049,0.080)  & 3.6 (1.5,5.5) & \textbf{179 (112,272)}                           &  \textbf{4.5 (3.1,6.3)}         \\ \hline
h(g) + SVGD + BD(b) & \textbf{0.050 (0.036,0.066)}  & 6.9 (2.9,11) & 243 (152,352)           
                &   4.9 (3.4,6.8 )        \\ \hline

\end{tabular}}
\caption{Anisotropic Gaussian. \textit{Rows:} SVGD, Birth-Death (BD), and Mixed Method. h(m) denotes bandwidth update using the median trick, while h(g) uses gradient descent. BD(b) indicates backtracking in Birth-Death. \textit{Columns:} KL divergence, KL after 10 iterations, total number of iterations, and running time.  Each numerical entry shows the mean followed by a 95\% confidence interval in parentheses. Bolded values represent the best or near-best entries in each column.}
\label{tb:AnisotropicGaussian}
\end{table}

\subsubsection{Anisotropic Gaussian Mixture with Unequal Weights}

In this section, we examine an anisotropic Gaussian with unequal weights. We set our target distribution to be $
\boldsymbol{y}_i \sim \frac{1}{5} \mathcal{N} \left(
\begin{bmatrix} -10 \\ 0 \end{bmatrix}, 
\begin{bmatrix} 5 & 0 \\ 0 & 0.5 \end{bmatrix}
\right) + \frac{4}{5} \mathcal{N} \left(
\begin{bmatrix} 10 \\ 0 \end{bmatrix}, 
\begin{bmatrix} 0.5 & 0 \\ 0 & 5 \end{bmatrix}
\right)$. We use $N = 100$ particles and the initial particles are sampled from with two-dimensional standard Gaussian distribution. This example is more challenging due to its strong anisotropy and unequal weighting, making it difficult for both the Birth-Death and SVGD methods when used separately. The results are presented in Table \ref{tb:anisotropic gaussian mixture with unequal weights}. As in the previous examples, updating bandwidth is a crucial step that promotes the convergence of all algorithms. The mixed method outperforms both SVGD and the Birth-Death method, as SVGD struggles with the limitation of transferring weights between two separated modes, while the Birth-Death method does not fully leverage the optimization properties of the SVGD algorithm.

\begin{table}[htbp]
\centering
\resizebox{\columnwidth}{!}{
\begin{tabular}{llllll}
\hline
                & KL & KL(10)    & \# iterations   & running time   \\ \hline\hhline{======}
 h(m) + SVGD      & diverge  & diverge  &  diverge                          &  diverge       \\ \hline
h(g) + SVGD      & 1.1 (0.92,1.3)  & 37 (32,40) &     330 (132,552)                        &    7.9 (4.2,11)     \\ \hline\hhline{======}

h(m) + BD        & diverge & diverge &     diverge               &    diverge   \\ \hline
h(g) + BD      &  0.44 (0.30,0.55)  & 5.5 (3.5,7.9) &          214 (82,382)                   &      6.2 (5.8,8.6)     \\ \hline

h(g) + BD(b)         &  0.40 (0.34,0.45) & 1.8 (1.0,2.9) &    \textbf{129 (52,262)}               &     \textbf{4.2 (2.7,6.7)}      \\  \hline\hhline{======}

h(m) + SVGD + BD       &  diverge & diverge &     diverge                         &   diverge     \\ \hline
h(g) + SVGD + BD   &   0.23 (0.19,0.30)&  4.6 (3.0,6.7) &                   167 (92,270)         &  14 (6.0,21)       \\ \hline
h(g) + SVGD + BD(b) & \textbf{0.19 (0.13,0.27)} & \textbf{1.3 (0.68,2.4)} &    263 (81,503)         
                &    7.8 (4.4,12.6)      \\ \hline

\end{tabular}}
\caption{Anisotropic Gaussian Mixture with Unequal Weights. \textit{Rows:} SVGD, Birth-Death (BD), and Mixed Method. h(m) denotes bandwidth update using the median trick, while h(g) uses gradient descent. BD(b) indicates backtracking in Birth-Death. \textit{Columns:} KL divergence, KL after 10 iterations, total number of iterations, and running time.  Each numerical entry shows the mean followed by a 95\% confidence interval in parentheses. Bolded values represent the best or near-best entries in each column.}
\label{tb:anisotropic gaussian mixture with unequal weights}
\end{table}

\subsection{Test Distribution}

\begin{figure}[htbp]
\begin{center}

   \begin{minipage}{0.32\textwidth}
     \centering
     \includegraphics[width=1\linewidth]{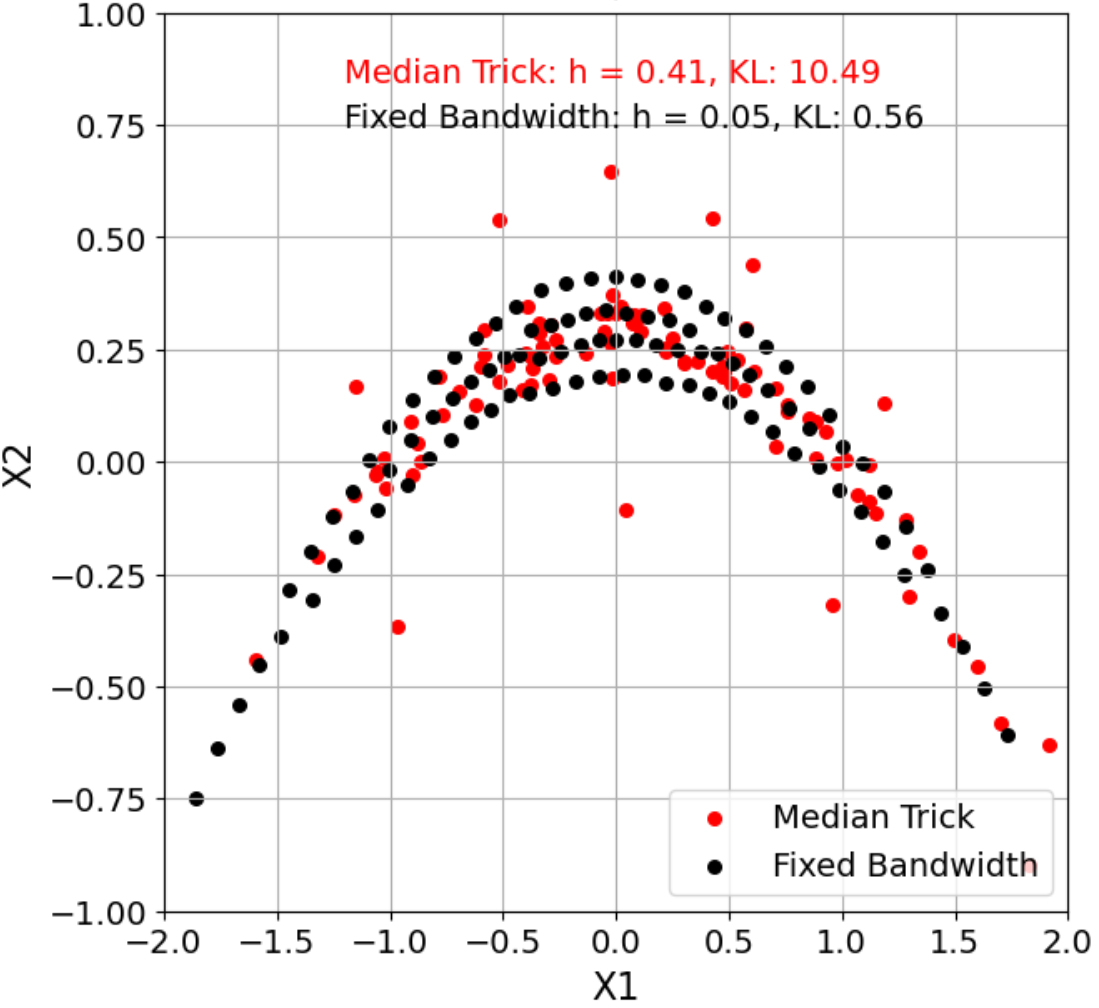}
   \end{minipage}
      \begin{minipage}{0.32\textwidth}
     \centering
     \includegraphics[width=1\linewidth]{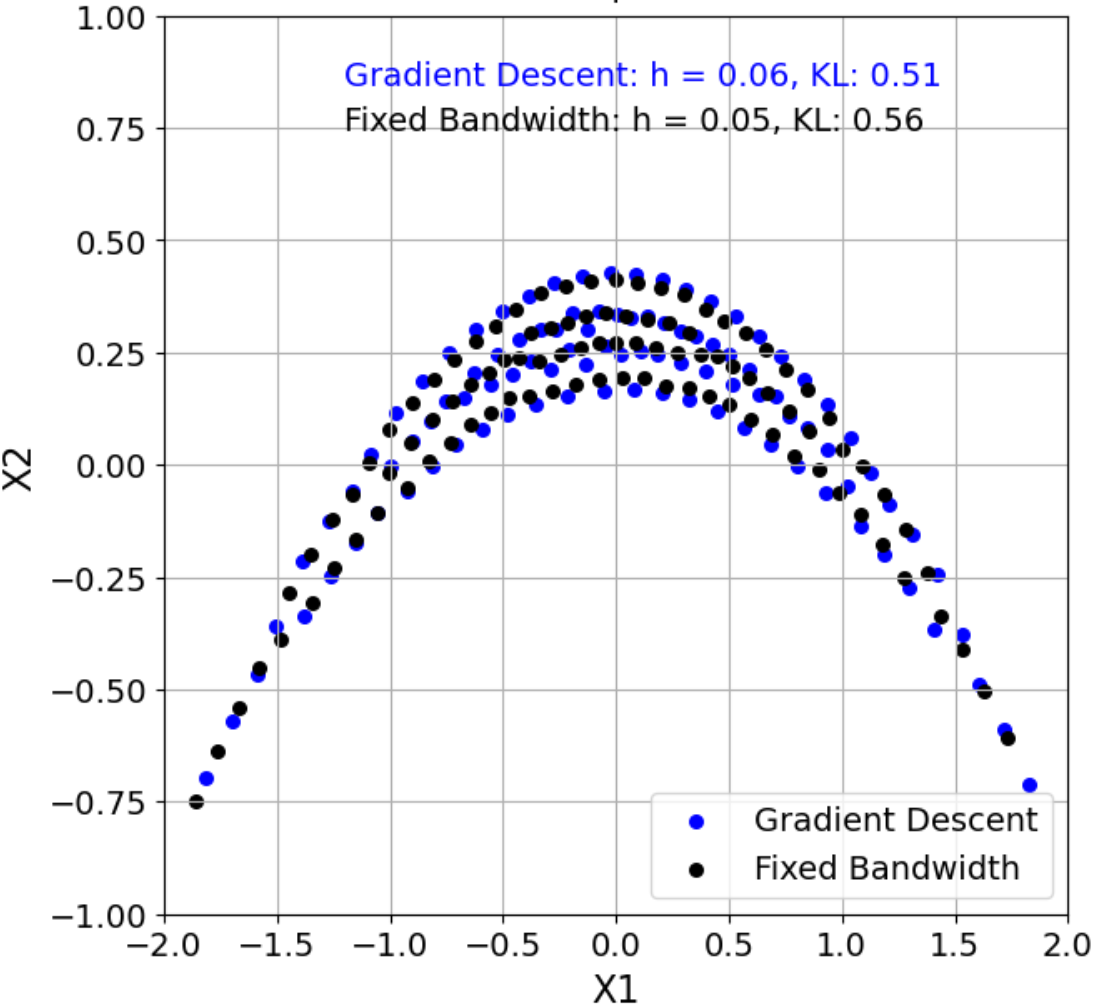}
   \end{minipage}
   \begin{minipage}{0.32\textwidth}
     \centering     \includegraphics[width=0.97\linewidth]{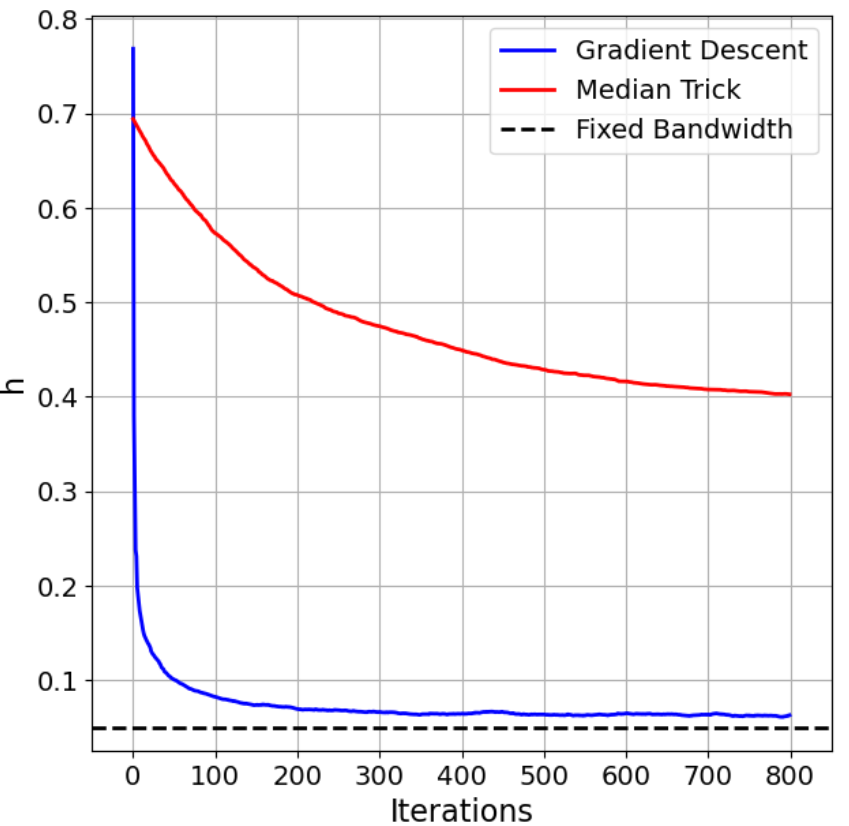}
   \end{minipage}
   
   \end{center}
   \caption{ \textit{Left:} Particles generated by SVGD with an adaptive bandwidth updated via the median trick heuristic. \textit{Middle:} Particles generated by mixed method with an adaptive bandwidth updated via gradient descent. \textit{Right:} The bandwidth over iterations. Note that the adaptive bandwidth converges near the fixed bandwidth value $h = 0.05$ and achieves the best KL-divergence.} 
   
   \label{fig:EVI}
\end{figure}

In this section, we test the example that is introduced in \cite{wang2021particle}. The target distribution is given by \[
\rho(\mathbf{x}) \propto \exp \left\{ -\frac{x_1^2}{2} - \frac{1}{2} \left(10x_2 + 3x_1^2 - 3 \right)^2 \right\}.
\]Figure \ref{fig:EVI} presents the particles returned by SVGD with a fixed bandwidth, SVGD with an adaptive bandwidth using the median trick, and the mixed method with an adaptive bandwidth updated via gradient descent. The fixed bandwidth value ($h = 0.05$), that is used in \cite{wang2021particle}, proves to be suitable for this particular case. However, the choice of bandwidth is not fully discussed. It is not clear how it could be chosen a priori.

Our adaptive bandwidth is initialized with the median of the pairwise distance as suggested in \cite{liu2016stein} and is updated with KL-based gradient descent. The value of bandwidth over the iterations is shown in the right panel of  Figure \ref{fig:EVI}. We observe that the adaptive bandwidth converges near the fixed bandwidth value and achieves a better KL-divergence than using a fixed bandwidth. Therefore, this suggests a way to avoid the need for carefully selecting a fixed bandwidth and instead dynamically updating it.  The median trick dramatically overestimates the best bandwidth, converging to 0.41 instead of 0.05, resulting in an overdispersed particle approximation, and a much larger KL divergence.

\begin{table}[h]
\centering
\resizebox{\columnwidth}{!}{
\begin{tabular}{llllll}
\hline
                & KL & KL(10)    & \# iterations   & running time   \\ \hline\hhline{======}
 h(0.05) + SVGD           & 0.55 (0.54,0.57)  & 8.7 (8.2,9.2) &\textbf{142 (52,262)}                       &0.83 (0.50,1.2)          \\ \hline
 h(m) + SVGD      &  2 (17,33)   &  8.7 (8.2,9.2) & 142 (52,262)                             &   0.79 (0.48,1.2)         \\ \hline
h(g) + SVGD      & \textbf{0.52 (0.51,0.53)}  & 41 (26,49) & 266 (112,382)                            &  1.2 (0.69,1.6)         \\ \hline\hhline{======}
h(0.05) + BD              & 0.88 (0.79,1)  & 2.1 (1.7,2.6) &288 (112,512)                              & 0.93 (0.50,1.5)          \\ \hline
h(m) + BD         & diverge & diverge&  diverge                     & diverge       \\ \hline
h(g) + BD         & 0.67 (0.62,0.72)  &3.0 (1.8,4.3)& 206 (92,382)                  &     \textbf{0.71 (0.43,1.2)}      \\  \hline
h(g) + BD(b)      &  0.81 (0.70,0.95)   & 3.5 (1.9,5.5) & 183 (72,342)                             & 0.90 (0.55,1.4)          \\ \hline\hhline{======}
h(0.05) + SVGD + BD       & 0.65 (0.62,0.67)   & \textbf{0.87 (0.76,0.99)} & 146 (32,322)                             & 0.88 (0.47,1.5)          \\ \hline
h(m) + SVGD + BD       &  diverge    &  diverge  &  diverge                              &  diverge          \\ \hline
h(g) + SVGD + BD   & 0.59 (0.56,0.62)  & 2.7 (1.4,3.9) & 228 (102,392)                            & 1.2 (0.73,1.8)         \\ \hline
h(g) + SVGD + BD(b) & \textbf{0.57 (0.53,0.62)}  & 3.4 (1.8,5.5) &  288 (131,492)            
                &   1.4 (0.84,2.1 )        \\ \hline

\end{tabular}}
\caption{Test distribution. \textit{Rows:} SVGD, Birth-Death (BD), and Mixed Method. h(m) denotes bandwidth update using the median trick, while h(g) uses gradient descent. BD(b) indicates backtracking in Birth-Death. \textit{Columns:} KL divergence, KL after 10 iterations, total number of iterations, and running time.  Each numerical entry shows the mean followed by a 95\% confidence interval in parentheses. Bolded values represent the best or near-best entries in each column.}
\end{table}
% \begin{figure}
% \begin{center}
%     \centering
% \includegraphics[width=.4\linewidth]{figure/EVI_SVGD+BD+h(g)_hs.png}
%    \end{center}
% \caption{The adaptive bandwidth over iterations. Note that the adaptive bandwidth converges to the fixed bandwidth value (h = 0.05).}
% \label{fig:hs}
% \end{figure}

\subsection{Exchangeable Multimodal Posteriors}
In this section, we consider an example of $n$-D Bayesian hierarchical model. We start with $n = 2$. We assume a linear observation model 
\begin{equation}
    b = Ax + \epsilon,  \quad \epsilon \sim \mathcal{N}\left(0, I\right)
\end{equation}
 where $A = [1,1]^{\top}$ and $b = 10$. 
 
 Since $A$ is rank one, it admits a one dimensional nullspace parallel to the vector $[1,-1]$. As a result, the likelihood alone can never distinguish the value of the difference $x_1 - x_2$. So, while the sum $x_1 + x_2$ is identifiable from the data, the difference $x_1 -x_2$ is not. This example is chosen to highlight the influence of the prior, which will act mainly to select candidate locations along the direction unresolved by the likelihood. This direction is intentionally symmetric with respect to relabeling, so that the posterior is always symmetric under reflection. Under this construction $x_1,x_2$ are exchangeable, so the posterior either admits one mode, centered at $x_1 = x_2$, or admits two modes separated along the $[1,-1]$ direction, and that are symmetric under reflection across the $x_1 = x_2$ line.

\begin{figure}[h!]
\begin{center}
   \begin{minipage}{0.32\textwidth}
     \centering
     \includegraphics[width=1.0\linewidth]{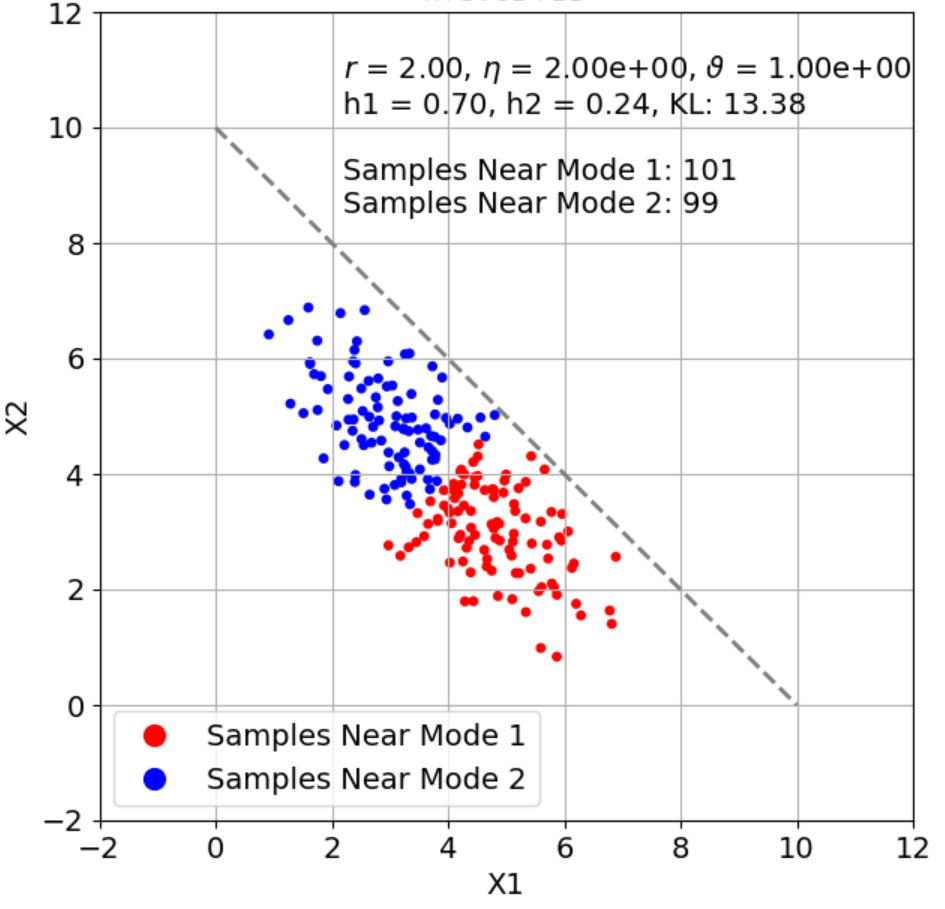}
   \end{minipage}
   \hspace{0.01cm}
   \begin{minipage}{0.32\textwidth}
     \centering
     \includegraphics[width=1.0\linewidth]{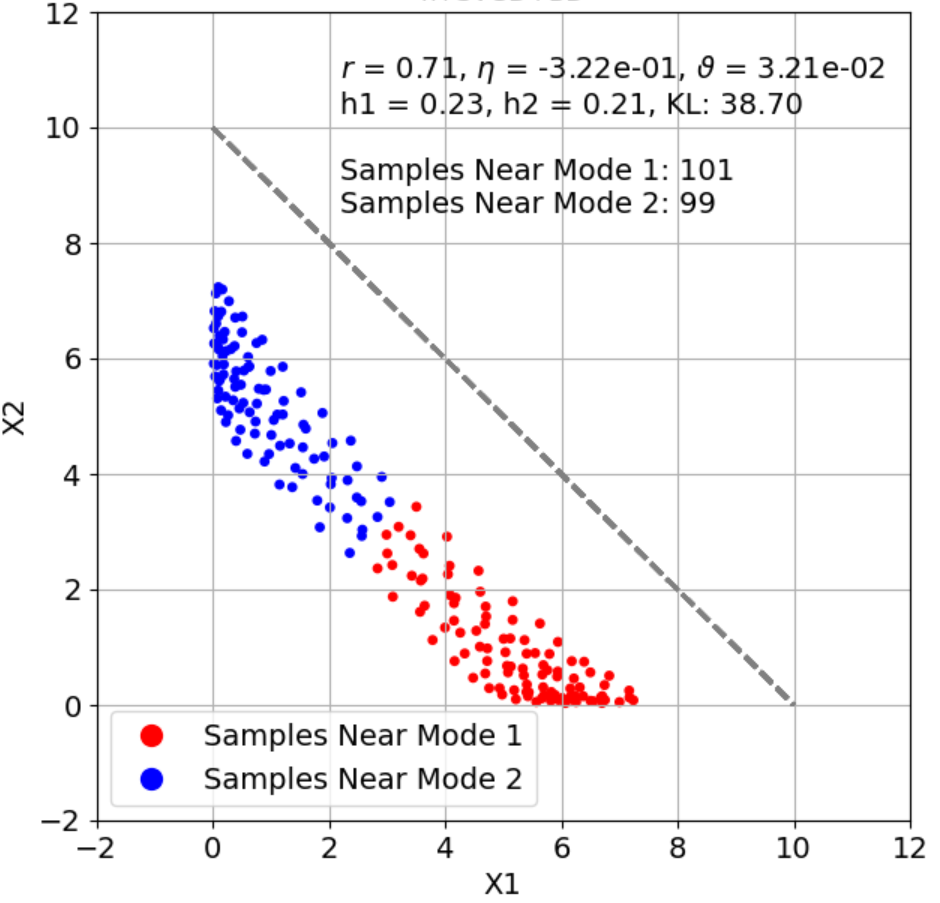}
   \end{minipage}
      \begin{minipage}{0.32\textwidth}
     \centering
     \includegraphics[width=1\linewidth]{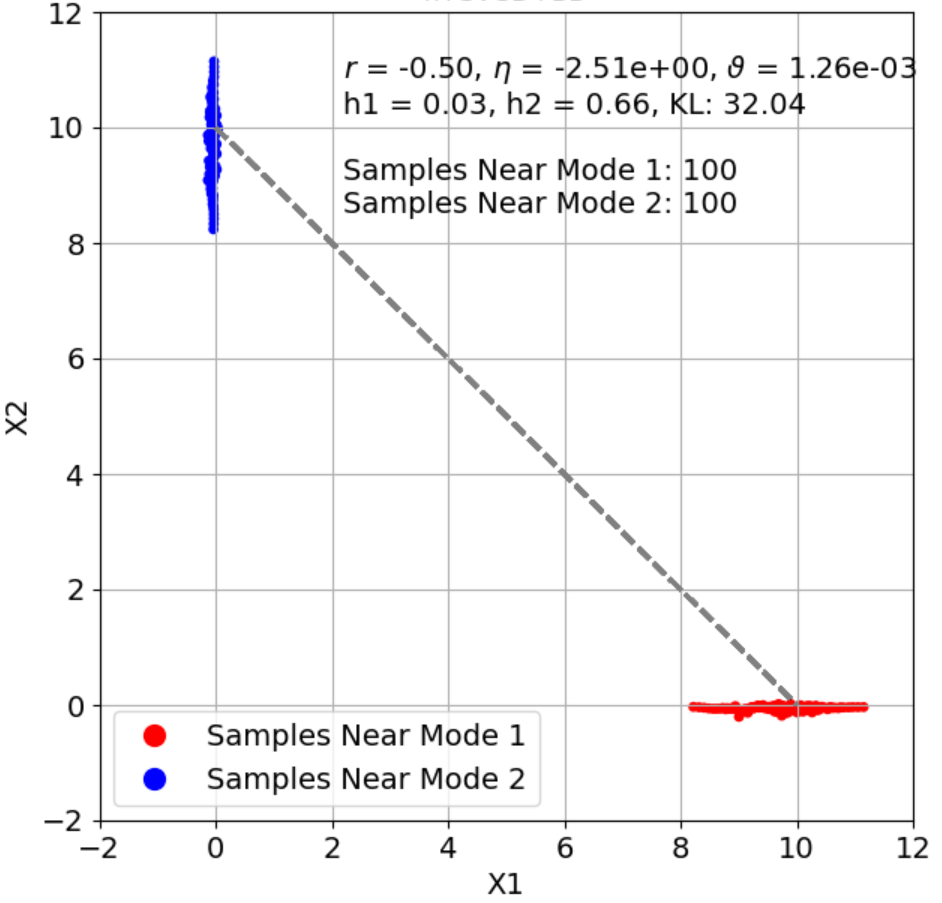}
   \end{minipage}
   \end{center}
      \caption{Particles at the starting (left), intermediate (middle), and ending points (right) of the parameter path. Note that the unimodal distribution splits into two separate modes as $r$ changes from $2$ to $-0.5$. The blue points represent samples near the upper left mode, while the red points represent samples near the lower right mode. The split occurs at around $r = 0.71$. The modes are well separated at the endpoint, $r = -0.5$. The dashed line $x + y = 10$ represents the MLE's without regularization.} 
   \label{fig:bayes}
\end{figure}

To test the continuation approach, we vary the hyperparameters along the line from a convex setting, $(r(0) , \eta(0), \vartheta(0)) = (2,2,1)$, to the desired non-convex setting $(-0.5,-2.5,10^{-4})$. For simplicity, we adopt 50 equidistant time points along the path. As discussed in \cite{calvetti2020sparsity,si2024path}, the prior promotes sparsity more strongly as $r(t)$ decreases. Therefore, the posterior distribution is expected to approach a bimodal distribution, with each mode located around $(0,10)$ and $(10,0)$. Figure \ref{fig:bayes} displays the particles corresponding to 3 different points on the hyperparameter path. At the starting point ($r = 2$), the distribution is uni-modal. As the parameter $r$ approaches 0.6, the distribution begins to split into two separated modes, and as $r$ approaches -0.5, it becomes entirely separated into two distinct modes. The number of particles within each mode is nearly the same, indicating the effectiveness of the birth-death process in directly maintaining the correct distribution of mass between modes.

\begin{table}[h]
\centering
\begin{tabular}{lllllll}
\toprule
dimension                & 2   & 4   & 8   & 16  & 32  & 64  \\ \midrule
total number of particles      & 100 & 100 & 200 & 200 & 400 & 400  \\ \midrule
minimum particle count in any mode &    50   & 23     &  23   &  10   &  10   &  3   \\ \midrule
observed CV &  0   &  0.06   & 0.08     &  0.15   &  0.16   &  0.22     \\ 
\bottomrule
\end{tabular}
\caption{High-dimensional multimodal posteriors. While mode discovery becomes increasingly challenging when the number of particles grows more slowly than the number of modes, the continuation methods successfully identify all modes even in high-dimensional settings.}
\label{tb:bayesHD}
\end{table}

To extend the model to $n$-dimensional cases, we set $ A = [1, \dots, 1]^{1\times n} $ and $ b = 10 $. In an $n$-dimensional setting, the model transitions from a unimodal distribution to a multimodal one with $n$ distinct, symmetrically distributed, modes as $r$ approaches zero. Detecting all modes and transferring mass between modes becomes increasingly challenging as $n$ grows for methods that do not use continuation, and do not assume strong problem symmetries. Directly applying SVGD is sensitive to initialization, as particles tend to cluster around the nearest modes, often failing to explore and capture all modes effectively. On the contrary, the continuation method, which starts from an easily sampled unimodal distribution and gradually transitions to the multimodal distribution, is more robust and more likely to discover all modes.
 The results are summarized in table \ref{tb:bayesHD}. Note that the continuation method succeeds in discovering all modes in every case tested, even with $n = 64$. 
 
 Since the unknowns are exhangeable under this model, all modes should have equal mass. The coefficient of variation (CV) in the number of particles assigned to each mode increased as the dimensionality grew, since the number of particles used grew more slowly than the number of modes. Using more particles reduced the CV in the mass assigned to the modes.

\subsection{High dimensional example}

In this section, we consider a high-dimensional Bayesian hierarchical model.  We test the  deconvolution problem introduced in \cite{calvetti2019hierachical}. This problem is selected since it is a standard test case (see also, \cite{calvetti2020sparse,calvetti2024computationally,si2024path}). 

Let $f:[0,1] \rightarrow \mathbb{R}$ be a piecewise constant function with $f(0)=0$. The data $y$ is generated by the following convolution:
$$
y_{j}=\int_{0}^{1} A\left(s_{j}-t\right) f(t) d t+\eta_{j}, \quad 1 \leq j \leq n, \quad A(t)=\left(\frac{J_{1}(\kappa|t|)}{\kappa|t|}\right)^{2}
$$
where $J_{1}$ is the Bessel function of the first kind, and $\kappa$ is a scalar controlling the width of the kernel. We set $\kappa=40$, $s_{j}=(4+j) / 100$. The above integral can be discretized as
$$
y=A v+\eta, \quad A_{j k}=w_{k} A\left(s_{j}-t_{k}\right), \quad \eta \sim \mathcal{N}\left(0, \gamma^{2} I_{n}\right),
$$
where $v \in \mathbb{R}^{d}$ has components $v_{k}=f\left(t_{k}\right)$ with $t_{k}=(k-1) /(n-1)$, and the $w_{k}$ are quadrature weights for discretization of the integral. We set the standard deviation $\gamma$ to be $1 \%$ of the max-norm of the noiseless signal.

Let $y_{j}=x_{j}-x_{j-1}$ with $x_{0}=0$. Since $x$ is piecewise constant, $y$ is sparse. 

$$
y=L x, \quad L = \left[\begin{array}{cccc}
1 & 0 & \ldots & 0 \\
-1 & 1 & \ldots & 0 \\
& & \ddots & \\
0 & \ldots & -1 & 1
\end{array}\right] \in \mathbb{R}^{n \times n}
$$

Then $x=\mathrm{C} y$ with
$$
C = L^{-1}=\left[\begin{array}{cccc}
1 & 0 & \ldots & 0 \\
1 & 1 & \ldots & 0 \\
\vdots & & \ddots & \\
1 & \ldots & 1 & 1
\end{array}\right] \in \mathbb{R}^{n \times n}
$$

The inverse problem is to estimate the assumed sparse vector $y$ from the noisy data vector $b$, given the forward operator $AC$.

$$
b=AC y+\epsilon, \quad \epsilon \sim \mathcal{N}\left(0, \sigma^{2} I\right), \quad a_{j k}=w_{k} A\left(s_{j}-t_{k}\right)
$$

As discussed in \cite{calvetti2020sparse}, the MAP estimates of vector $y$ is sensitive to the choice of hyperparameters. The continuation method traces the solution path as the hyperparameters vary, thereby aiding in the selection of suitable hyperparameters. With this idea, a path-following method for MAP estimates is proposed in \cite{si2024path}. Similarly, for sampling, the continuation method tracks the posterior samples as the hyperparameters change. 

The hyperparameter 
$r$ controls the sparsity promoting effect of the model \cite{si2024path}. As 
$r$ decreases, the model promotes sparsity more strongly, leading the posterior distribution to transition from a unimodal to a multimodal distribution. Directly sampling from this posterior can be challenging due to the ill-posed nature of Bayesian linear problems. To address this, a Markov Chain Monte Carlo (MCMC) based method is proposed in \cite{calvetti2024computationally}. However, this method requires initialization with MAP estimates. When $r<1$, computing MAP estimates involves solving a non-convex ill-posed optimization problem, which has been shown to be numerically difficult in \cite{calvetti2020sparsity}. In addition, it only considers three specific values of $r$---namely, $r = 1, -0.5$ and $-1$. As a result, it remains unclear how the samples evolve as the hyperparameters vary.  In particular, the case $r = -1$ presents additional sampling challenges due to strong correlations between $x$ and $\theta$, which induce a posterior that is tightly concentrated along a  a narrow, curved manifold. This structure can significantly hinder the efficiency of standard MCMC methods, leading to poor mixing and low acceptance rates.

\begin{figure}[t!]
\begin{center}
   \begin{minipage}{0.32\textwidth}
     \centering
     \includegraphics[width=1.0\linewidth]{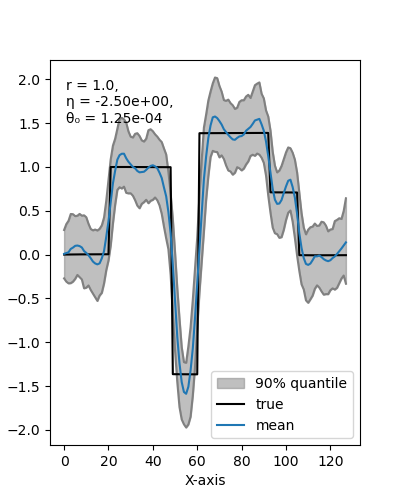}
   \end{minipage}
   \hspace{0.01cm}
   \begin{minipage}{0.32\textwidth}
     \centering
     \includegraphics[width=1.0\linewidth]{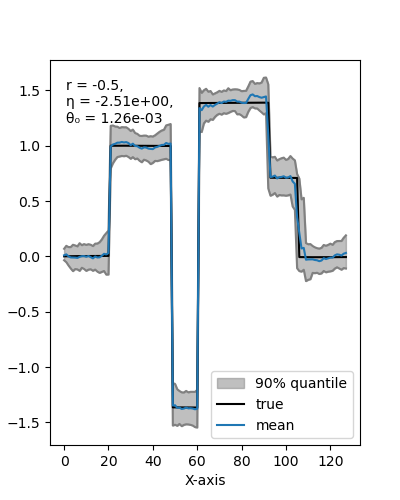}
   \end{minipage}
   \begin{minipage}{0.32\textwidth}
     \centering
     \includegraphics[width=1\linewidth]{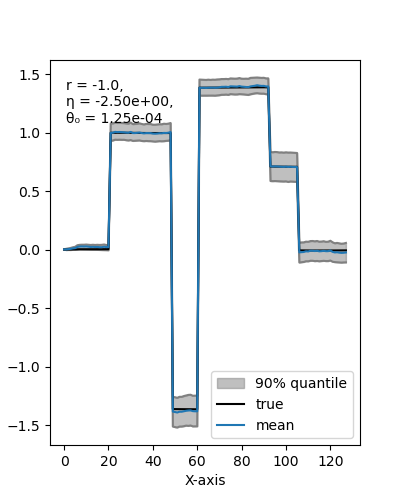}
   \end{minipage}
   
   \vspace{0.3cm} % Adds spacing between rows

   \begin{minipage}{0.32\textwidth}
     \centering
     \includegraphics[width=1.0\linewidth]{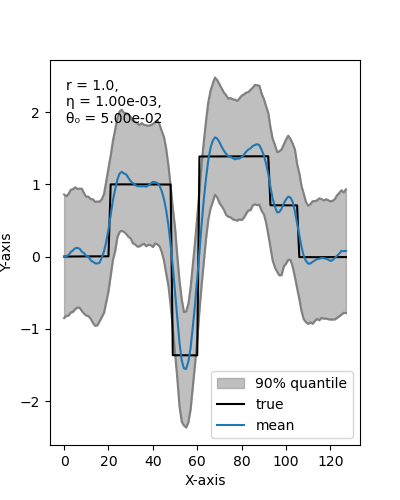}
   \end{minipage}
   \hspace{0.01cm}
   \begin{minipage}{0.32\textwidth}
     \centering
     \includegraphics[width=1.0\linewidth]{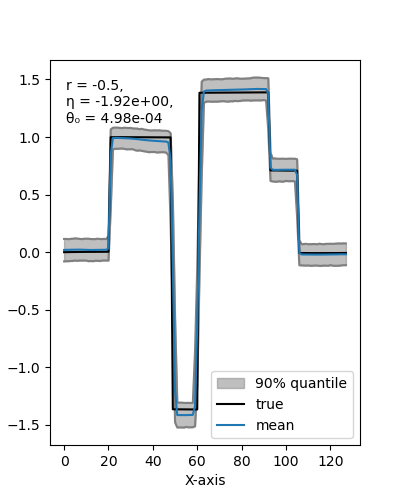}
   \end{minipage}
   \begin{minipage}{0.32\textwidth}
     \centering
     \includegraphics[width=1\linewidth]{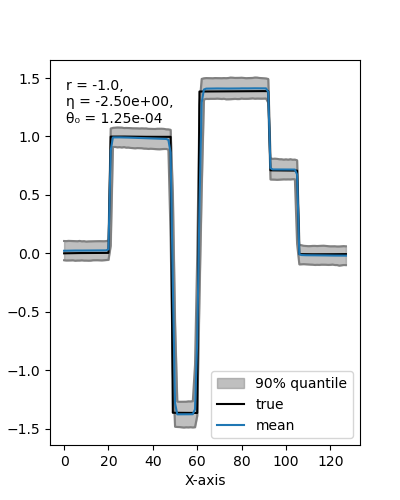}
   \end{minipage}
\end{center}
\caption{\textit{Top row:} Particles obtained using MCMC initialized with MAP estimates for $r = 1$, $r = -0.5$, and $r = -1$. 
\textit{Bottom row:} Particles at the starting (left), intermediate (middle), and ending points (right). Note that the sparsity-promoting effect becomes stronger as $r$ decreases, indicating a transition from a unimodal distribution to a multimodal distribution.} 
\label{fig:combined}
\end{figure}

In contrast, the continuation method allows us to start from a unimodal distribution ($r\geq 1$) , which is easier to sample from, and gradually follow a chosen hyperparameter path to reach the target multimodal distribution ($r<1$). For comparison, we use $r = 1$ as the starting point and $r = -1$ as the ending point, using 40 equally spaced steps along the path. Figure \ref{fig:combined} displays the particles at the starting, intermediate ($r = -0.5) $, and the ending points. We observe that the particles at the endpoint closely match those obtained using the MCMC method, indicating the effectiveness of the continuation method. In all cases shown, our posterior estimates are smoother than those produced by direct sampling, and exhibit less ``ringing" in the posterior mean. While the posterior intervals are roughly equivalent, we will show that the continuation method is far more successful at discovering posterior modes.

Additionally, we note that the continuation method is more robust, as it does not rely on MAP estimate initialization, which is crucial for the success of the MCMC method. Furthermore, this approach provides the entire particle path, allowing us to select a suitable $r$ value and obtain the desired particles.

\subsubsection{Mode discovery}

In this section, we compare the results by MCMC method \cite{calvetti2024computationally} and continuation method in terms of mode discovery. The target distribution is the posterior at $r = -0.5$, which is a strongly multimodal distribution. Since the MCMC particles are all initialized with MAP estimates, they are prone to insufficient sample space exploration. In contrast, the continuation is more likely to explore multiple modes effectively. 

To determine the number of modes discovered, we perform clustering on both sets of particles using the same threshold for the minimum distance between clusters. The results show that the MCMC method detects 12 modes, while the continuation method detects 40 modes. Next, we combine the particles and compute the proportion of total particles captured by the modes identified by MCMC and the continuation method, respectively. With a maximum distance to the centroids set at 2.5, MCMC modes cover only 58.25\% of the total particles, while continuation modes cover 96.35\% of the total particles.

\subsubsection{Coverage Check}

Since the previous examples used a fixed true signal, it is not possible to directly analyze the accuracy of the approximate posterior except by comparison to gold standard posterior estimators. Since no such procedure exists for the highly multimodal examples considered, we could not verify the approximated posteriors against a ground truth.

In this section, we evaluate the continuation method based on the actual coverage rate of the confidence interval from the generated models. By generating many sample problems from the prior model, we can evaluate the actual accuracy of posterior approximations produced by our continuation method. First, we sample the variance \( \theta \) from a generalized Gamma distribution with hyperparameters \( (r, \beta, \theta_0) \). Then, we sample signal \( x \) from a Gaussian distribution with variance \( \theta \). The observation vector \( y \) is generated using the same forward model $A$,  

\[
y = A x + \epsilon,
\]
where $\epsilon$ is the standard Gaussian noise.

To evaluate the effectiveness of the continuation method, we compute confidence intervals for the magnitude of several selected dimensions and the posterior log-likelihood using the obtained particles. We conduct 1,000 generated trials, counting the number of times the true value falls within the corresponding confidence interval. We then compare the actual coverage rate to the expected quantiles of the confidence interval. The results are presented in Table \ref{tab:coverage_rates} and \ref{tab:coverage_rates_loglikelihood}. Notably, the coverage rates for the magnitude across all chosen dimensions and the likelihood closely match the true quantiles of their respective confidence intervals.

\begin{table}[t!]
    \centering
    \begin{tabular}{r c c c c c c}
        \toprule
        & \textbf{20} & \textbf{40} & \textbf{60} & \textbf{80} & \textbf{100} & \textbf{120} \\
        \midrule
        \textbf{95\%}  & 0.945 & 0.958 & 0.942 & 0.940 & 0.937 & 0.953 \\
        \textbf{90\%}  & 0.911 & 0.903 & 0.903 & 0.911 & 0.903 & 0.911 \\
        \textbf{80\%}  & 0.816 & 0.822 & 0.798 & 0.827 & 0.814 & 0.837 \\
        \textbf{60\%}  & 0.604 & 0.617 & 0.614 & 0.614 & 0.643 & 0.638 \\
        \textbf{40\%}  & 0.438 & 0.423 & 0.430 & 0.415 & 0.433 & 0.420 \\
        \textbf{20\%}  & 0.231 & 0.218 & 0.210 & 0.207 & 0.244 & 0.215 \\
        \bottomrule
    \end{tabular}
    \caption{Coverage rates at different quantile levels for the chosen dimensions}
    \label{tab:coverage_rates}
\end{table}

\begin{table}[t!]
    \centering
    \begin{tabular}{r c c c c c c}
        \toprule
         & \textbf{95\%} & \textbf{90\%} & \textbf{80\%} & \textbf{60\%} & \textbf{40\%} & \textbf{20\%} \\
        \midrule
        \textbf{10\%}  & 0.938 & 0.865 & 0.833 & 0.604 & 0.365 & 0.198 	 \\
        \textbf{90\%}  &  	0.927& 0.906 	 & 0.740 & 0.594 	 &  	0.375 	 & 0.156 \\
        \bottomrule
    \end{tabular}
    \caption{Coverage rates at different quantile levels for the selected dimensions, where the dimensions correspond to the 10\% and 90\% quintiles of the variance $\theta$.}
    \label{tab:coverage_rates}
\end{table}

\begin{table}[h]
    \centering
    \begin{tabular}{lcccccc}
        \toprule
         & \textbf{95\%} & \textbf{90\%} & \textbf{80\%} & \textbf{60\%} & \textbf{40\%} & \textbf{20\%} \\
        \midrule
        \textbf{coverage} & 0.938 & 0.906 & 0.812 & 0.688 & 0.458 & 0.198 \\
        \bottomrule
    \end{tabular}
    \caption{Coverage rates at different quantile levels for the likelihood}
    \label{tab:coverage_rates_loglikelihood}
\end{table}

\appendix

\section{Gradient with respect to the bandwidth}\label{appx:bd}
In this section, we provide a detailed derivation of the gradient used to update the bandwidth $h$. 

We start with an one-dimensional case. We retain the same notation, where $\rho^{*}$ denotes the target distribution. To simplify the variational density, we omit the explicit dependence on $h$, writing $\rho = \rho_{h}$ as: 
$$
\rho = \rho_{h} = \frac{1}{N} \sum_{i=1}^{N} \frac{1}{\sqrt{2\pi} h} \exp{\frac{-(x-x_{i})^{2}}{2h^{2}}},
$$
where $x_{i}$ are samples. Let:
\begin{equation}
    \kappa(x,x';h) = \frac{1}{\sqrt{2 \pi} h} \exp\left( -\frac{1}{2} \frac{(x - x')^2}{h^2} \right)
\end{equation}
denote the Gaussian kernel which is also dependent on the bandwidth $h$.

Notice that 
$$
\frac{\partial D_{KL}(\rho \parallel \rho^{*})}{\partial h} = \frac{\delta D_{KL}(\rho \parallel \rho^{*})}{\delta \rho} \frac{d\rho}{dh},
$$
where
\begin{equation}
\begin{aligned}
\frac{\delta D_{KL}(\rho \parallel \rho^{*})}{\delta \rho} = &\frac{d D_{KL}(\rho + \varepsilon \phi \parallel \rho^{*})}{d\varepsilon}\bigg|_{\varepsilon=0} \\
= & \frac{d}{d\varepsilon} \int \ln{\frac{\rho + \varepsilon \phi}{\rho^{*}}}(\rho + \varepsilon \phi) dx \bigg|_{\varepsilon=0} \\ 
= & \int \ln{\frac{\rho + \varepsilon \phi}{\rho^{*}}} \phi + \frac{\rho^{*}}{\rho + \varepsilon \phi}(\rho + \varepsilon \phi) \frac{\phi}{\rho^{*}}dx  \bigg|_{\varepsilon=0} \\
= & \int (\ln{\frac{\rho}{\rho^{*}}} + 1)\phi
\end{aligned}
\end{equation}

Here $\phi$ is the direction $\frac{d\rho}{dh}$, and $\int \phi = \frac{d}{dh} \int \rho = 0$.

If we only know the density up to a constant $C$, the result is the same, i.e
\begin{equation}
        \frac{\delta D_{KL}(\rho \parallel \rho^{*})}{\delta \rho} 
        = \int \ln{\frac{\rho}{\rho^{*}}}\phi
        =  \int (\ln{\frac{\rho}{\rho^{*}}} - \ln C)\phi
        = \int \ln{\frac{\rho}{C \rho^{*}}}\phi.
\end{equation}

Then,  $\frac{\partial D_{KL}(\rho \parallel \rho^{*})}{\partial h} = \int \ln{\frac{\rho}{\rho^{*}}} \frac{d\rho}{dh} dx$.

By plugging in $\rho = \frac{1}{N} \sum_{i=1}^{N} \frac{1}{\sqrt{2\pi} h} \exp{\frac{-(x-x_{i})^{2}}{2h^{2}}}$,
\begin{equation}
\begin{aligned}
\frac{\partial D_{KL}(\rho \parallel \rho^{*})}{\partial h} & = \int \ln{\frac{\rho}{\rho^{*}}} \frac{1}{N} \sum_{i=1}^{N} \frac{1}{h} \left( \frac{(x-x_{i})^{2}}{h^{2}} - 1 \right) \frac{1}{\sqrt{2\pi} h} \exp{\frac{-(x-x_{i})^{2}}{2h^{2}}} \\
& = \int \ln{\frac{\rho}{\rho^{*}}} \frac{1}{h} \frac{1}{N} \sum_{i=1}^{N} \left( \frac{(x-x_{i})^{2}}{h^{2}} - 1 \right) \kappa(x,x_i;h)
\end{aligned}
\end{equation}

For $d$-dimensional case, suppose covariance matrix $\Sigma = h^{2}I$, then 
\begin{align*}
\rho
&  = \frac{1}{N} \sum_{i=1}^{N} \frac{1}{\sqrt{(2\pi)^{d}\det(\Sigma)}} \exp(\frac{-(x-x_{i})^{\intercal} \Sigma^{-1}(x-x_{i})}{2}) \\
& = \frac{1}{N} \sum_{i=1}^{N} \frac{1}{\sqrt{(2\pi)^{d}}h^{d}} \exp(\frac{- \|x-x_{i}\|_{2}^{2}}{2 h^{2}}).
\end{align*}
Then after plugging in,
\begin{equation}
\begin{aligned}
\frac{\partial D_{KL}(\rho \parallel \rho^{*})}{\partial h} 
&= 
\int \ln{\frac{\rho}{\rho^{*}}}
\frac{1}{N} \sum_{i=1}^{N} 
\frac{1}{h} \left( \frac{\|x-x_{i}\|_{2}^{2}}{h^{2}} - d \right)
\frac{1}{\sqrt{(2\pi)^{d}}h^{d}} \exp(\frac{- \|x-x_{i}\|_{2}^{2}}{2 h^{2}}) \\
&=
\int \ln{\frac{\rho}{\rho^{*}}} \frac{1}{h} \frac{1}{N} \sum_{i=1}^{N} \left( \frac{\|x-x_{i}\|_{2}^{2}}{h^{2}} - d \right) \kappa(x,x_i;h) .
\end{aligned}
\end{equation}

\bibliographystyle{plain}
\bibliography{ref}
\end{document}